\documentclass[%
 reprint,
 amsmath,amssymb,
 aps,
showkeys
]{revtex4-2}

\usepackage{graphicx}
\usepackage{dcolumn}
\usepackage{bm}
\usepackage{hyperref}

\usepackage{hyperref}
\hypersetup{colorlinks, linkcolor=blue}

\begin{document}


\title{Critical-point phenomena and finite-size scaling in mean-field \mbox{equal-coupling photonic networks}}

\author{Oliver Melchert}
 \email{melchert@iqo.uni-hannover.de}
\affiliation{%
Leibniz Universität Hannover, Institute of Quantum Optics, Welfengarten 1, 30167 Hannover, Germany}
\affiliation{
Cluster of Excellence PhoenixD (Photonics, Optics, and Engineering - Innovation Across Disciplines), Welfengarten 1, 30167 Hannover, Germany
}%

\date{\today}

\begin{abstract}
%
The mean-field optical phase transition in multimode equal-coupling photonic networks is studied by temporal evolution of the nonlinear equations of motion of the coupled modes.
Analogies to statistical mechanics models of interacting classical spins, built upon the correspondence between complex-valued modes and two-component spins, are employed to define two-component and single-component order parameters.
A comprehensive finite-size scaling analysis is performed to estimate critical points and exponents of a second-order phase transition, driven by the optical energy per mode. 
Equilibrium properties of the system are compared to exact results whenever applicable.
Considering various parameter settings, our results confirm the mean-field nature of the transition and establish the critical line in the nonlinearity--energy-density phase diagram. 
Critical scaling leads to infer the upper critical dimension $d_c=4$.
Connection to thermodynamic quantities is established by means of a kinetic temperature with appropriate zero-temperature limit. In the low temperature phase (low energy per mode), spins align. In the high temperature phase (large energy per mode), spins rotate independent of one another.
\end{abstract}

                              
\maketitle


\section{\label{sec:intro} Introduction}

In statistical mechanics it is usually the goal to describe system behavior in the thermodynamic limit, i.e.\ under the condition that the size of an elementary building block 
of a system is miniscule in comparison to the overall size of the system \cite{Cardy:BOOK:1996,Newman:BOOK:1999,Binder:BOOK:2010}. 
Much effort is thus spend on designing algorithms and numerical strategies that permit to efficiently simulate systems with a large but finite number $N$ of building blocks \cite{Hartmann:BOOK:2001,Newman:BOOK:1999,Binder:BOOK:2010}, and to reliably extrapolate towards $N\to\infty$, e.g., employing the heuristical statistical mechanics theory of finite-size scaling (FSS)  \cite{Binder:ZPB:1981,Stanley:RMP:1999,Cardy:BOOK:1996,Stanley:BOOK:1987}.
In case of photonic networks, however, systems of interest naturally exhibit a finite number of building blocks, and effects caused by their finite size may even become an important feature.
This is especially important for the study of collective phenomena that require a system-wide cooperation of building blocks, such as phase transitions.

Based on these considerations we study fully connected, equal-coupling photonic networks (ECPNs), consisting of $N$ complex-valued modes that are subject to nonlinear effects and interact pairwise via uniform couplings.
%
%
For such long-range interacting ECPNs, recent studies provided evidence for a second-order phase transition \cite{Ramos:PRX:2020}.  
This optical phase transition can be expressed in terms of statistical mechanics spin-model terminology, if a modes phase is interpreted as a two-component spin vector, and its amplitude as the length of that vector. This geometric picture exploits analogies to models of interacting classical spins \cite{Cardy:BOOK:1996,Binder:BOOK:2010,Stanley:BOOK:1987}, as, e.g., the XY model \cite{Stanley:PRL:1968,Tobochnik:PRB:1979,XY:NOTE,Kim:PRE:2001,Hong:PRE:2015}.
The analogy of light propagating through nonlinear two-dimensional (2D) arrays of weakly coupled waveguides and the 2D short-range interacting XY model has previously been used to study an optical analog of the Berezinskii-Kosterlitz-Thouless transition \cite{Small:PRA:2011}.
On long-range interacting systems, the XY model exhibits a temperature driven, continuous phase transition with a ferromagnetic phase at low temperature and a paramagnetic phase at high temperature \cite{Kim:PRE:2001,Hong:PRE:2015}. 
In case of the fully connected ECPN model, working in the microcanonical ensemble, the transition is observed as a function of the energy density per mode.
If the energy per mode is large, the system is in a disordered state in which the angular degrees of freedom of the spins are active and where there is no orientational order amongst the spins.
If the average energy in each mode decreases, the system undergoes a second order phase transition to an ordered phase \cite{Ramos:PRX:2020}, in which alignment forces induce global order. 
%
%
Previously, Ref.\ \cite{Ramos:PRX:2020} demonstrated the role of the nonlinearity in achieving this optical phase transition and reported the critical point and verified the mean-field scaling of a magnetization-like order parameter for a particular parameter setting. 

Here, we perform a more detailed analysis, based on numerical simulations of ECPNs of finite size, employing FSS to thoroughly characterize the optical phase transition in terms of critical points and critical exponents. 
%
%
Since fully connected ECPNs represent a long-range interacting system, the transition can be expected to be in the mean-field universality class, consistent with the mean-field theory for photonic networks developed in Ref.~\cite{Ramos:PRX:2020}. 
Due to the inevitably finite number of building blocks of real photonic systems, a thorough study of finite-size effects in such systems seems rewarding.

In Sec.~\ref{sec:model} we introduce equal-coupling photonic networks in more detail,
and we introduce classical statistical mechanics quantities that allow to characterize the
state of the system.
Section~\ref{sec:results} contains a detailed analysis of equilibrium features of the model and of the optical phase transition, focusing on a parameter range considered previously in Ref.~\cite{Ramos:PRX:2020}. 
We first discuss emergence and loss of order in the system (Sec.~\ref{sec:res_order}), demonstrating that a symmetry of the Hamiltonian enables coherent phase-rotations in the ordered phase, and we derive exact results for time-averaged quantities in the disordered phase, even accounting for finite-size effects. 
Subsequently we employ FSS to precisely locate the critical point and determine the critical exponents from information on systems of finite size (Sec.~\ref{sec:res_fss}). 
Correlation times (Sec.~\ref{sec:res_ACtimes}), and scaling at criticality (Sec.~\ref{sec:res_scaling_hc}) are discussed separately.
In Sec.~\ref{sec:res_temp} we utilize a kinetic Temperature based on the angular momenta of the spins to analyze the thermodynamic signature of the transition within the considered dynamical system.
Finally, we summarize results obtained for various other parameter settings in terms of a phase diagram and make conclusions in Sec.~\ref{sec:d_and_c}.

\section{\label{sec:model} Model and Methods}

Fully connected, equal coupling photonic networks, consisting of $N$ complex-valued modes $\Psi \equiv (\psi_1,\ldots, \psi_N)$ that interact pairwise via uniform couplings $J_0>0$, are described by the Hamiltonian \cite{Ramos:PRX:2020}
\begin{equation}
\mathcal{H}[\Psi] = -\frac{J_0}{N} \sum_{\ell,j\neq \ell} \psi_\ell^* \psi_j + \frac{\chi}{2} \sum_\ell |\psi_\ell|^4,\label{eq:H}
\end{equation}
where the second term models a Kerr-type nonlinearity with nonlinearity parameter $\chi>0$.
%
%
The total optical power of a mode configuration $\Psi$
reads
\begin{equation}
\mathcal{A}[\Psi] = \sum_{\ell} |\psi_\ell|^2, \label{eq:A}
\end{equation}
and the time evolution of the system (\ref{eq:H}) is governed by the nonlinear equations of motion 
\begin{eqnarray}
i\dot{\psi}_\ell = -\frac{J_0}{N} \sum_{j\neq \ell} \psi_j + \chi |\psi_\ell|^2
\psi_\ell,\label{eq:EOM}
\end{eqnarray}
i.e.\ $N$ coupled first-order differential equations where the dot denotes derivative with respect to time. They
conserve the energy density $h\equiv \mathcal{H}[{\Psi}]/N$ and the optical power per mode $a\equiv\mathcal{A}[\Psi]/N$. 
The norms $|\psi_\ell|$ of the individual modes are not conserved.
The equations of motion (\ref{eq:EOM}) are obtained in the framework of time-dependent coupled mode theory \cite{Ramos:PRX:2020,Shi:PRR:2021}.
Equation~(\ref{eq:H}) is invariant under the transformation $\psi_\ell \to \psi_\ell \exp\{i\theta\}$ for $\ell=1,\ldots,N$, indicating a continuous symmetry that causes degeneracy of distinct states: a global phase-rotation $\theta$ will conserve the energy density $h$.
This symmetry is also manifested in the dynamics of the system, see Sec.~\ref{sec:res_order}.
Instances $\Psi$, evolving according to the equations of motion Eq.~(\ref{eq:EOM}), approach equilibrium dynamics for $h$ in the range from $h_{\rm min}=-J_0 a + \chi a^2/2$ to $h_{\rm{max}}= \chi a^2$ \cite{Ramos:PRX:2020}.
In the limit of weak nonlinearity it is directly possible to test whether a system exhibits equilibrium dynamics: the framework of optical thermodynamics facilitates comparison of the average optical power within the different eigenmodes of the system to theoretical predictions \cite{Ramos:PRX:2020,Wu:NP:2019,Wu:CP:2020,Parto:OL:2019}.
Moreover, in weakly nonlinear systems, the thermalization process can be controlled by dispersion engineering \cite{Shi:PRR:2021}.
Values of $\chi$ considered in this work do not qualify as ``weak'', hence we will need to use a different strategy to assess whether a simulation run has reached dynamical equilibrium, see Sec.~\ref{sec:res_ACtimes}.

%
In analogy to statistical mechanics models of interacting $n$-dimensional classical spins  \cite{Stanley:PRL:1968,Cardy:BOOK:1996}, and, more specifically, in analogy to the XY model (obtained for $n=2$) \cite{Stanley:PRL:1968,Tobochnik:PRB:1979,Kim:PRE:2001,Hong:PRE:2015}, the modes $\psi_\ell$ can be viewed as \emph{photonic soft-spins}
\begin{equation}
\mathbf{s}_\ell=(s_{x,\ell},s_{y,\ell})\equiv (\mathsf{Re}[\psi_\ell], \mathsf{Im}[\psi_\ell]), \label{eq:s}
\end{equation}
vectors of variable length in the $\mathsf{Re}[\psi]$-$\mathsf{Im}[\psi]$--plane.
The \emph{softness} of a spin, i.e.\ the variable length of the spin vector, is a feature that distinguishes photonic networks from classical spin models \cite{XY:NOTE}. 
Inspired by such statistical mechanics models we introduce the two-component magnetization-like \emph{order parameter} \cite{Cardy:BOOK:1996,Yamaguchi:PRE:2019,Kim:PRE:2001,Leoncini:PRE:1998,Newman:BOOK:1999}
\begin{equation}
\mathbf{m}=(m_x, m_y)= \frac{1}{N} \sum_\ell \mathbf{s}_\ell,\label{eq:mag}
\end{equation}
to analyze the soft-spins collective dynamics.
Let us note that for the fully connected photonic network, Eq.~(\ref{eq:EOM}) can be expressed in terms of ${\mathbf{s}}_\ell$ and ${\mathbf{m}}$
as
\begin{equation}
i\dot{\mathbf{s}}_{\ell} = - J_0\,\left(\mathbf{m} - \frac{\mathbf{s}_\ell}{N}\right) + \chi |\mathbf{s}_\ell|^2\mathbf{s}_{\ell},  \label{eq:EOM_s}
\end{equation}
wherein $J_0\mathbf{m}$ serves as a mean field guiding the time evolution of individual spins, and where the second term couples the components of the spin vector.
%
To investigate emergence of order in the ECPN, we further consider the single-component
order parameter $m\equiv |\mathbf{m}|$. Its long-time average 
\begin{align}
\langle m \rangle = \frac{1}{t_{\mathrm{obs}}-t_{\mathrm{eq}}} \int_{t_{\mathrm{eq}}}^{t_{\mathrm{obs}}}~m(t)~{\mathrm{d}}t, \label{eq:m}
\end{align}
with equilibration time $t_{\mathrm{eq}}$ and total observation time $t_{\mathrm{obs}}$,
allows to analyze equilibrium properties as function of the \emph{control parameter} $h$, and to probe the nature of the optical phase transitions supported by Eq.~(\ref{eq:H}). 
While the quantity $m$ has also been used in a previous study \cite{Ramos:PRX:2020}, we here show that the two-component order parameter ${\mathbf{m}}$ plays a leading role.
It permits much deeper insight into the dynamics, allowing to identify collective rotational modes, and to derive exact results for several quantities in the limit $h\to h_{\mathrm{max}}$.
%

%
Subsequently, we fix the optical power per mode to $a=1$ and consider parameter values $J_0=1.2$ (fixing the value of $J_0$ is equivalent to fixing time units), and $\chi=1$. The equilibration range is bounded by $h_{\mathrm{min}}=-0.7$ and $h_{\mathrm{max}}=1$.  This allows to reproduce and built on previous results presented in Ref.~\cite{Ramos:PRX:2020}.
Numerical simulations for other parameter choices are pointed out explicitly in Sec.~\ref{sec:d_and_c}.
In the presented study, the equations of motion Eq.~(\ref{eq:EOM}) are solved for configurations $\Psi$ of up to $N=384$ modes, for long observation times up to $t_{\mathrm{obs}}=10^7$ time units. 
Numerical integration of Eq.~(\ref{eq:EOM}) is performed using a step size controlled  Runge-Kutta method of high order \cite{Hairer:BOOK:1993,pyecpn:GH:2022,DOP853:NOTE}. 
Working in the microcanonical ensemble, the (constant) energy density and optical power of the system is set by the initial condition.
In order to prepare intitial mode-configurations for specified optical power per mode $a$ and energy density $h$, we utilize an effective optimization heuristic that incrementally improves a random trial configuration, see Supplemental Material \cite{SuppMat}. 
The initial $t_{\rm{eq}}=10^6$ time units are reserved for the system to approach equilibrium. This  equilibration phase is much longer than the correlation times for even the largest system, see Sec.~\ref{sec:res_ACtimes}. No measurements are taken during that time.

\begin{video}[tb!]
\href{https://figshare.com/articles/media/Equilibrium_dynamics_on_ECPNs/19237185}{\includegraphics{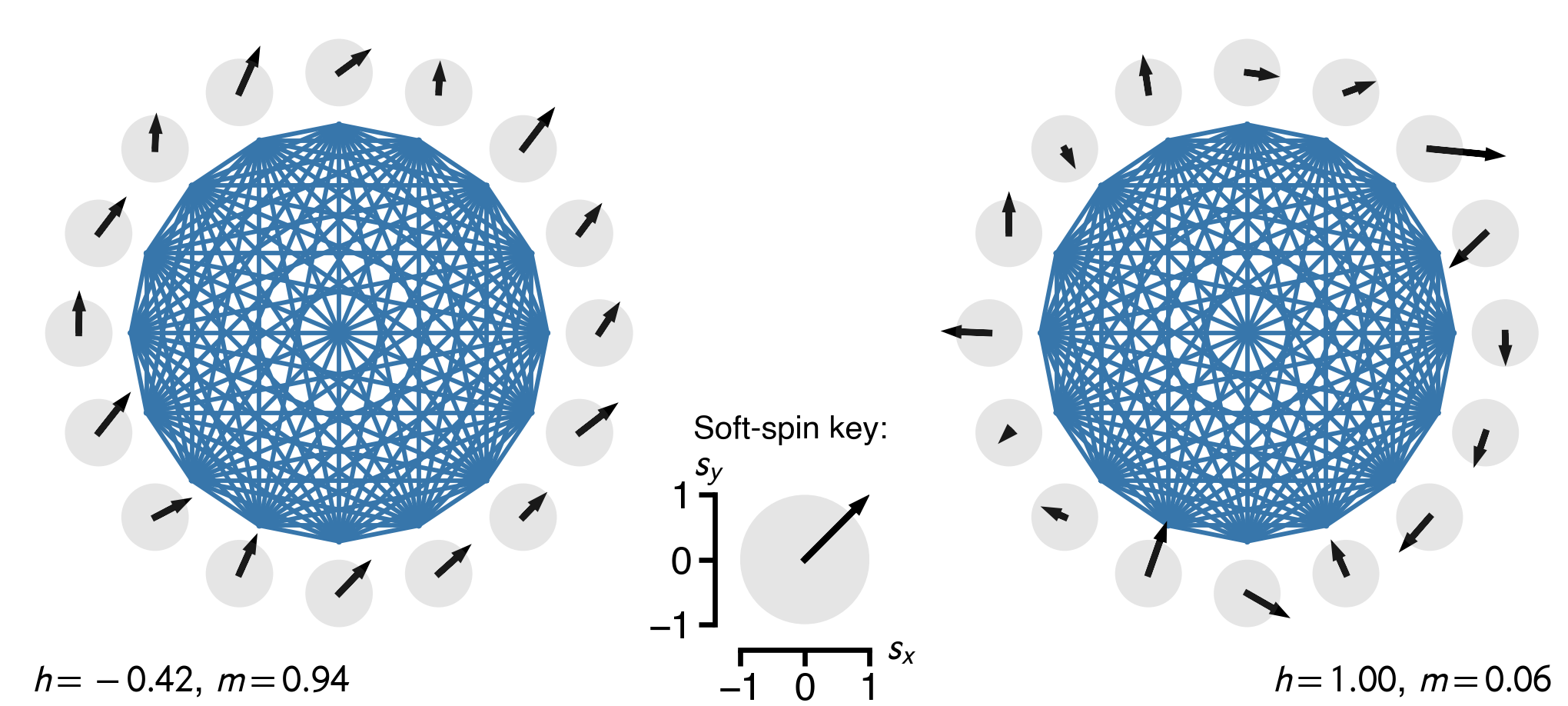}}
 \caption{\label{vid:1} 
 Equilibrium dynamics of fully connected equal-coupling photonic networks. Configurations of $N=16$ soft-spins, demonstrating order and disorder at $h=-0.42$ (left), and $h=1$ (right), respectively. Clip shows temporal evolution of both configurations for an interval of $1000$ time units with increment $\Delta t = 10$ between frames. 
}%
\end{video}

\section{\label{sec:results} Results}

Observing the time evolution of mode preparations $\Psi$ at different values of the control parameter $h$, allows to identify phases with distinct dynamics. 
Exemplary equilibrium dynamics, giving an account of order and disorder within an ensemble of $N=16$ soft spins, are shown in Vid.~\ref{vid:1}.
Both phases can be distinguished via the long-time average $\langle m \rangle$ of the order parameter.  
For $J_0=1.2$ and $\chi=1$, a previous qualitative analysis suggested a threshold value $h_c\approx 0.75$, separating an ordered phase with nonzero spontaneous magnetization ($h<h_c$), from a disordered phase  with an asymptotically ($N\to \infty$) zero value of $\langle m \rangle$ ($h>h_c$) \cite{Ramos:PRX:2020}.
%
Below, we derive exact results for several quantities in the limit $h \to h_{\mathrm{max}}$, and perform a thorough FSS analysis, locating the asymptotic critical point, and giving the critical exponents of the transition.

\begin{figure}[tb!]
\centering\includegraphics[width=\linewidth]{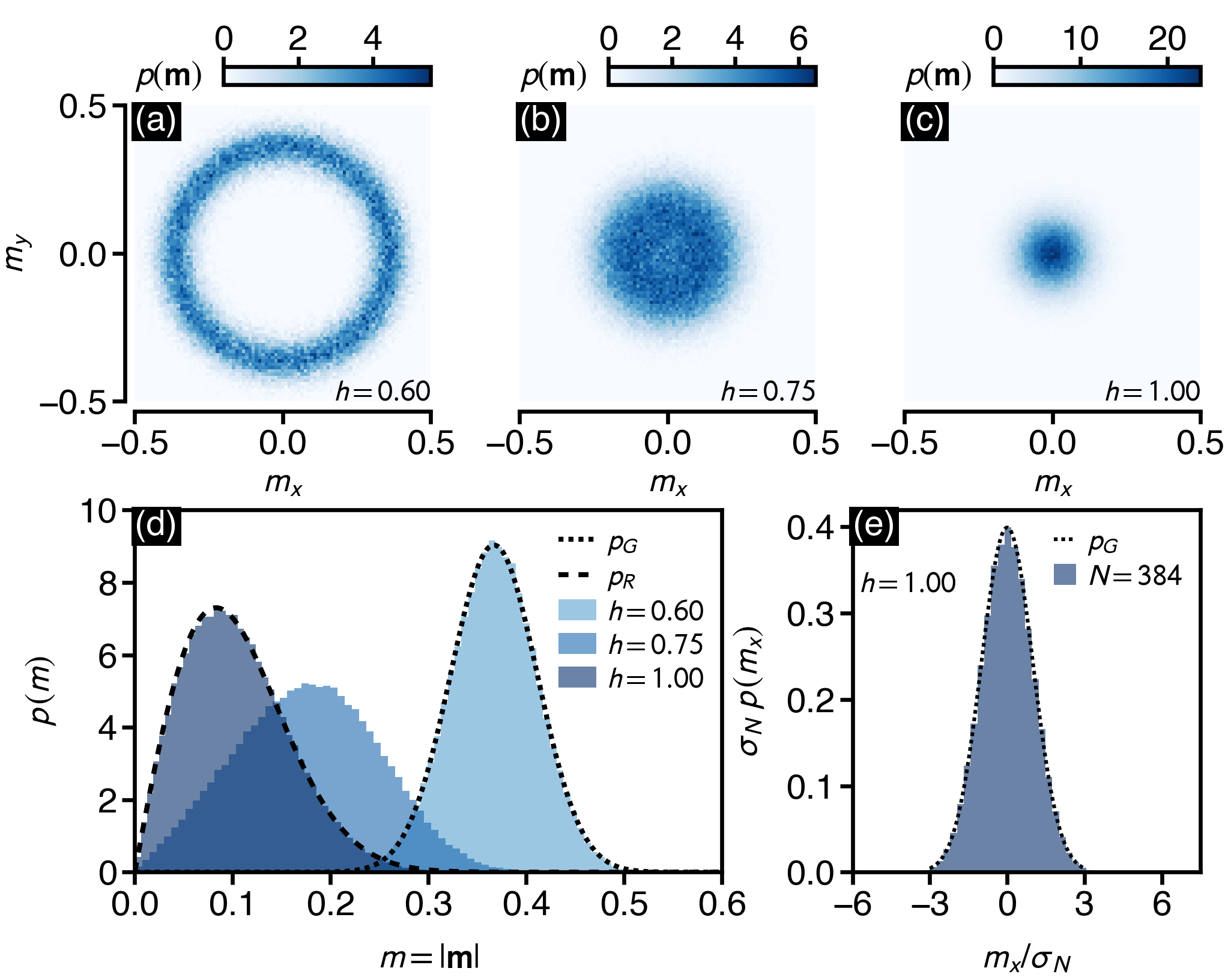}
\caption{
Distribution of the two-component order parameter ${\mathbf{m}}$ for $N=64$ at (a) $h=0.6$, (b) $h=0.75$, and, (c) $h=1.0$.
(d) Distribution of $m=|{\mathbf{m}}|$ corresponding to (a-c). $p_G$ is a Gaussian  with mean $\mu=0.367$ and variance $\sigma^2=0.002$. $p_R$ is a Rayleigh distribution with scale parameter $\sigma_R=0.083$. 
(e) Normalized distribution of only the $x$-component of ${\mathbf{m}}$ at $h=1$ for $N=384$. $p_G$ is a Gaussian  with $\mu=0$ and $\sigma^2=1$.
}
\label{fig:1}
\end{figure}

\subsection{\label{sec:res_order} Angular excitations and exact results}

In the ordered phase, as $h\to h_{\mathrm{min}}$, the spins align and loose their softness.
This becomes evident right at $h_{\mathrm{min}}$, where in absence of any length-fluctuations $|{\mathbf{s}}_\ell| = 1$ \cite{SuppMat}. 
%
In this limit, the spins equation of motion [Eq.~(\ref{eq:EOM_s})] take the approximate form
$i \dot{{\mathbf{s}}}_\ell \approx - J_0 {\mathbf{m}} + \chi{\mathbf{s}}_\ell$, where the right-hand side guides individual spins towards their average direction.  
%
Time-evolution proceeds mostly by directional changes [see left panel of Vid.~\ref{vid:1}], and it is safe to assume $\dot{\mathbf{s}}_\ell \approx |\mathbf{s}_\ell|\dot{\theta}_\ell\, {\mathbf{e}}_{\theta,\ell}$, with polar angle $\theta_\ell$ and unit vector ${\mathbf{e}}_{\theta,\ell}\equiv -\sin(\theta_\ell)\,{\mathbf{e}}_{x} + \cos(\theta_\ell)\,{\mathbf{e}}_{y}$.
For the entire system it follows that $\dot{\mathbf{m}} = (1/N)\sum_\ell \dot{\mathbf{s}}_\ell \approx m \dot{\theta}\, {\mathbf{e}}_{\theta}$, where $\theta$ is the polar angle of ${\mathbf{m}}$.
Since Eq.~(\ref{eq:H}) is invariant with respect to global phase rotation, the momentum corresponding to $\theta$ is conserved. Hence, we expect the ensemble of spins to be able to sustain collective rotational modes with constant angular frequency $\dot{\theta}$, allowing the system to move between distinct equilibrium configurations at fixed $h$.
%
This can be seen in Vid.~\ref{vid:1} for $h=-0.42$, where
$\dot{\theta}\approx -0.05$ ($\theta$ is measured in counterclockwise direction). Consequently, ${\mathbf{m}}$ completes a full rotation every $2\pi / |\dot{\theta}| \approx 125$ time units.  
These angular excitations represent the Goldstone modes of this classical long-range interacting system \cite{Cardy:BOOK:1996,Goldstone:PR:1962,Filho:PRE:2020}.
Such coherent rotations can be suppressed by preparing the system in initial states comprised of real-valued modes \cite{Ramos:PRX:2020}.
The angular momenta of the soft-spins relative to the collective rotational mode can be used as a measure of ``thermal'' excitation of the spins angular degrees of freedom, see Sec.~\ref{sec:res_temp}.
%
Let us note that the emergence of a nonzero magnetization within the ordered phase is signaled by both, ${\mathbf{m}}$ [Fig.~\ref{fig:1}(a-c)], and $m$ [Fig.~\ref{fig:1}(d)]. 
The distribution of ${\mathbf{m}}$ exhibits a circular shape with radius $m$.
In particular, at $h=0.6$, the distribution of $m$ compares well to a Gaussian with nonzero mean $\langle m \rangle \approx 0.37$ [Fig.~\ref{fig:1}(d)].
This spontaneous magnetization indicates the breakdown of the inherent isotropy, exhibited by equilibrium configurations within the disordered phase, see Vid.~\ref{vid:1} for $h=1$. 

%


%
Deep in the disordered phase, as $h\to h_{\rm{max}}$, the components $s_x$ and $s_y$ of the individual soft-spins have the properties of i.i.d.\ (independent and identically distributed) normal random variables with zero mean and variance $1/2$.
(Practically any random mode configuration, obtained using the simple sampling approach discussed in the Supplemental Material \cite{SuppMat}, will have these properties. Note that the distribution of energy densities obtained via simple sampling is also strongly peaked at $h=h_{\mathrm{max}}$.) 
Denoting a normal distribution with mean $\mu$ and variance $\sigma^2$ by $\mathcal{N}(\mu,\sigma^2)$, it is $s_{x,\ell} \sim \mathcal{N}(0,1/2)$ [and  $s_{y,\ell} \sim \mathcal{N}(0,1/2)$] for $\ell = 1,\ldots,N$. 
The above assumption was justified {\emph{a posteriori}} by performing numerical tests, see Supplemental Material \cite{SuppMat}.
As a result, $m_x\sim \frac{1}{N}\mathcal{N}(0,N/2)$ [and $m_y \sim \frac{1}{N}\mathcal{N}(0,N/2)$]. Thus, for given system size $N$, the time-average of the  components of ${\mathbf{m}}$ yield $\langle m_x\rangle \approx 0$ and $\langle m_y \rangle \approx 0$. 
At a given instant of time, however, $m_x$ and $m_y$ are normal i.i.d.\ random variables
with variance $\sigma^2 = 1/(2N)$ [since $c\, \mathcal{N}(\mu,\sigma^2)= \mathcal{N}(\mu,c^2\sigma^2)$]. Both features are demonstrated in Fig.~\ref{fig:1}(e), where the probability density of $m_x$ for a system of size $N=384$ is shown to fall onto a Gaussian function with zero mean and unit variance, if scaled by the standard deviation $\sigma_N=1/\sqrt{2N}$. 
%
As a result, ${\mathbf{m}}$ is a two-component normal random vector and Eq.~(\ref{eq:EOM_s}) can be cast into the form
$i\dot{\mathbf{s}}_\ell = {\mathbf{m}}_{{\mathrm{rand}},\ell} + \chi |{\mathbf{s}}_\ell|^2 {\mathbf{s}}_\ell$,
describing an ensemble of effectively uncoupled nonlinear oscillators, driven by fast-varying local random fields ${\mathbf{m}}_{{\mathrm{rand}},\ell} = -J_0 ({\mathbf{m}}-{\mathbf{s}}_\ell/N)$. 
The normal random nature of ${\mathbf{m}}$ has direct implications for $m$:
it specifies the length of a two-dimensional random vector, thus following a Rayleigh-distribution $p_R(m) = m \sigma_R^{-2}\exp\{-m^2/(2\sigma_R^2)\}$ with scale parameter $\sigma_R=1/\sqrt{2N}$ \cite{Papoulis:BOOK:1984}, see Fig.~\ref{fig:1}(d).
%
%
From this, the $n$-th moment about the origin $\mu_n^\prime \equiv \int m^n p_R(m)~{\mathrm{d}}m$ of the order parameter distribution  can be found explicitly \cite{Papoulis:BOOK:1984}: it is
 $ \mu_1^\prime = \sqrt{\pi/4/N}$, 
 $\mu_2^\prime = 1/N$, and
 $\mu_4^\prime = 2/N^2$.
%
With these results, the time-average of $m$ in the limit $h\to h_{\mathrm{max}}$ is found as $\langle m \rangle_N^\star = \mu_1^\prime \approx 0.89/\sqrt{N}$.
These exact limiting behaviors are tested in Sec.~\ref{sec:res_fss}, and Sec.~\ref{sec:res_scaling_hc} below.
%
Having discussed the role of the magnetization-like parameter ${\mathbf{m}}$ in the limits $h\to h_{\mathrm{min}}$ and $h\to h_{\mathrm{max}}$, it is now in order to study how systems composed of a large but finite number of photonic soft-spins gradually loose orientational order upon increasing the energy density $h$.  

\subsection{\label{sec:res_ACtimes} Correlation times}

\begin{figure}[tb!]
\centering\includegraphics[width=\linewidth]{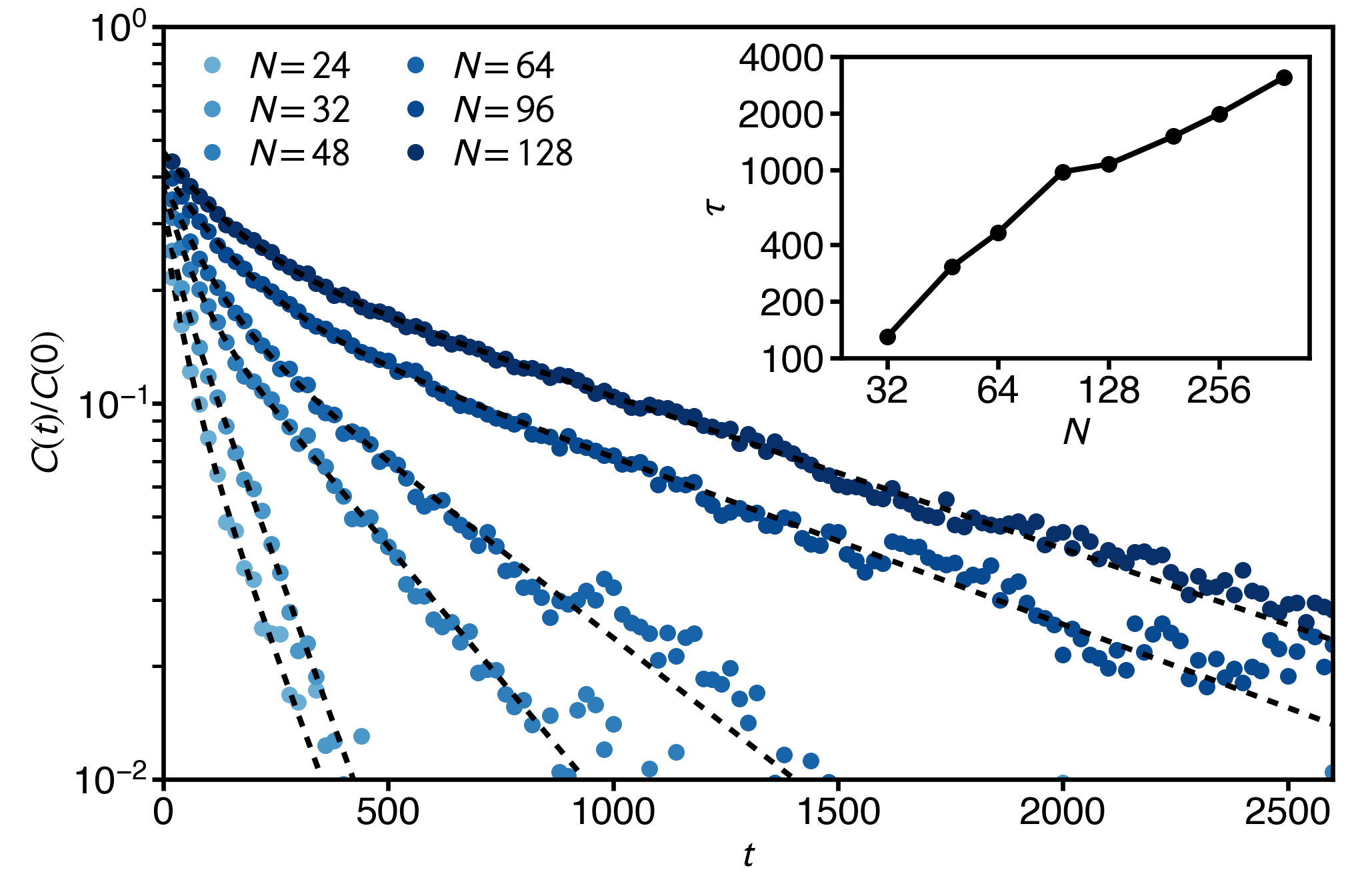}
\caption{Dependence of the autocorrelation function $C(t)$ on time-displacement $t$ for different number of modes $N$ at $h=0.75 \approx h_c$. Inset shows autocorrelation times $\tau$ obtained by least-squares fits of $C$ to a parameterized decay function (see text for details).}
\label{fig:C}
\end{figure}

To assess under which conditions a given simulation run allows to obtain equilibrium 
properties of $m$, we study the
time-displaced autocorrelation function \cite{Newman:BOOK:1999}
\begin{equation}
C(t) = \int_{t>t_{\mathrm{eq}}} \left[ m(t^\prime)m(t^\prime+ t) - \langle m \rangle^2 \right]~{\mathrm{d}}t^\prime. \label{eq:C}
\end{equation}
It takes a non-zero value if, on average, the fluctuations of $m(t^\prime)$ and its time-displaced value $m(t^\prime+t)$ are correlated. It will be zero if they are uncorrelated.
The analysis below is restricted to a control parameter in the vicinity of the critical point, where correlations are notoriously persistent \cite{Newman:BOOK:1999}. Specifically, we consider the point $h=0.75$, motivated by a previous study \cite{Ramos:PRX:2020}.
Figure~\ref{fig:C} shows autocorrelation functions for different system sizes $N$. 
The time-scale $\tau$ on which the autocorrelation decays is a measure of the correlation time, indicating how long it takes for small disturbances in $m$ to relax \cite{Binder:BOOK:2010}.
We find a good fit to the two-exponential decay function $C_{\mathrm{fit}}(t) \propto \exp(-t/\tau)\left[ 1+b \exp(-t/\tau_{\mathrm{corr}})\right]$ \cite{Wansleben:JAP:1987,Binder:BOOK:2010}. 
Results for $\tau$, obtained by least-squares fits to $C_{\mathrm{fit}}$, are shown in the inset of Fig.~\ref{fig:C}. 
We assume that if a system has evolved for several multiples of the correlation time, it is practically in equilibrium.
Irrespective of $N$, both, the initial equilibration time $t_{\mathrm{eq}}$ as well as the total observation time $t_{\mathrm{obs}}$ exceed the equilibration time by far.

\subsection{\label{sec:res_fss} Finite-size scaling in the critical region}
To investigate the emergence of ``magnetic ordering'' through soft-spin alignment, equilibrium properties of ECPNs as a function of the control parameter $h$ for systems of different size $N$ are considered.
%
%
As evident from Fig.~\ref{fig:3}(a), and in agreement with earlier results \cite{Ramos:PRX:2020}, an asymptotically nonzero magnetization appears below $h_c\approx 0.75$, and the transition from the disordered phase ($h>h_c$) to the ordered phase ($h<h_c$) becomes more clear as the size $N$ of the system increases. 
This is supported by the scaling behavior of other standard statistics mechanics quantities, such as the finite-size susceptibility $\chi_m$, and the Binder fourth-order cumulant $b_m$, given by 
\begin{subequations}
\begin{align}
\chi_m &\equiv N \left(\langle m^2\rangle - \langle m\rangle^2 \right), \label{eq:chi_m} \\
b_m &\equiv 1- \frac{\langle m^4\rangle}{3 \langle m^2\rangle^2}, \label{eq:b_m}
\end{align}
\end{subequations}
shown in Figs.~\ref{fig:3}(b,c), respectively.
FSS theory asserts \cite{Binder:ZPB:1981,Fisher:PRL:1972,Botet:PRL:1982,Botet:PRB:1983,Cardy:BOOK:1996}, that the scaling behavior of the above quantities satisfies the scaling forms
\begin{subequations}\label{eq:scaling}
\begin{align}
\langle m \rangle &= N^{-\beta/\bar{\nu}} f_\beta(x),\quad &f_\beta(x)&\stackrel{x\to 0}{\sim} (-x)^\beta, \label{eq:scaling_m}\\
\chi_m &= N^{\gamma/\bar{\nu}} f_\gamma(x), \quad &f_\gamma(x)&\stackrel{x\to0}{\sim}x^{-\gamma}, \label{eq:scaling_chi}\\
b_m &= f_0(x), \quad &f_0(x)&\stackrel{x\to 0}{\sim} {\mathrm{const.}}, \label{eq:scaling_b}
\end{align}
\end{subequations}
with scaling variable $x\equiv (h/h_c-1) N^{1/\bar{\nu}}$. 
$\bar{\nu}$ is a critical exponent governing the
divergence $N_c \sim |h-h_c|^{-\bar{\nu}}$ of a coherence number $N_c$ for the fully connected system in the limit $N\to\infty$ \cite{Botet:PRL:1982}.
The exponent $\bar{\nu}$ is related to the mean-field critical exponent $\nu$ through the relation $\bar{\nu}=d_c\nu$ \cite{Romano:PRE:2014,Ellis:AAP:2010}, where $d_c$ is the upper critical dimension \cite{Cardy:BOOK:1996}, i.e.\ the dimension above which short-range interacting systems have the same critical exponents as the  mean-field model.
$\beta$ is a critical exponent describing the vanishing of the order parameter $\langle m \rangle$, and
$\gamma$ describes the divergence of $\chi_m$ at the transition point. 
The scaling functions $f_\beta$, $f_\gamma$, and $f_0$ are expected to approach constant values for $x\to \infty$.
The scaling assumptions Eqs.~(\ref{eq:scaling}) are valid in the limit $N\to\infty$ with systematic deviations for systems of finite size \cite{Binder:BOOK:2010}. Consequently, the FSS analysis below is performed for systems of size $N \geq 96$.
Different methods for performing FSS are used throughout the literature, based, e.g., on the minimization of a local linearity function \cite{Kawashima:JPSJ:1993}, a refined quality function \cite{Houdayer:PRB:2004,Melchert:PRB:2009,autoScale:GH:2022}, considered in the presented work, and Bayesian inference methods \cite{Harada:PRE:2011,Harada:PRE:2015}.
Since the Hamiltonian Eq.~(\ref{eq:H}) describes a long-range interacting system, we expect our results to be in accord with the usual mean field exponents $\nu=1/2$, $\beta=1/2$, and $\gamma=1$ \cite{Cardy:BOOK:1996,Stanley:BOOK:1987}. In particular, the relation $\bar{\nu}=d_c/2$ will allow to estimate the upper critical dimension of ECPNs.

\subsubsection{\label{sec:res_fss_b}Binder fourth order cumulant}
\begin{figure}[tb!]
\centering\includegraphics[width=\linewidth]{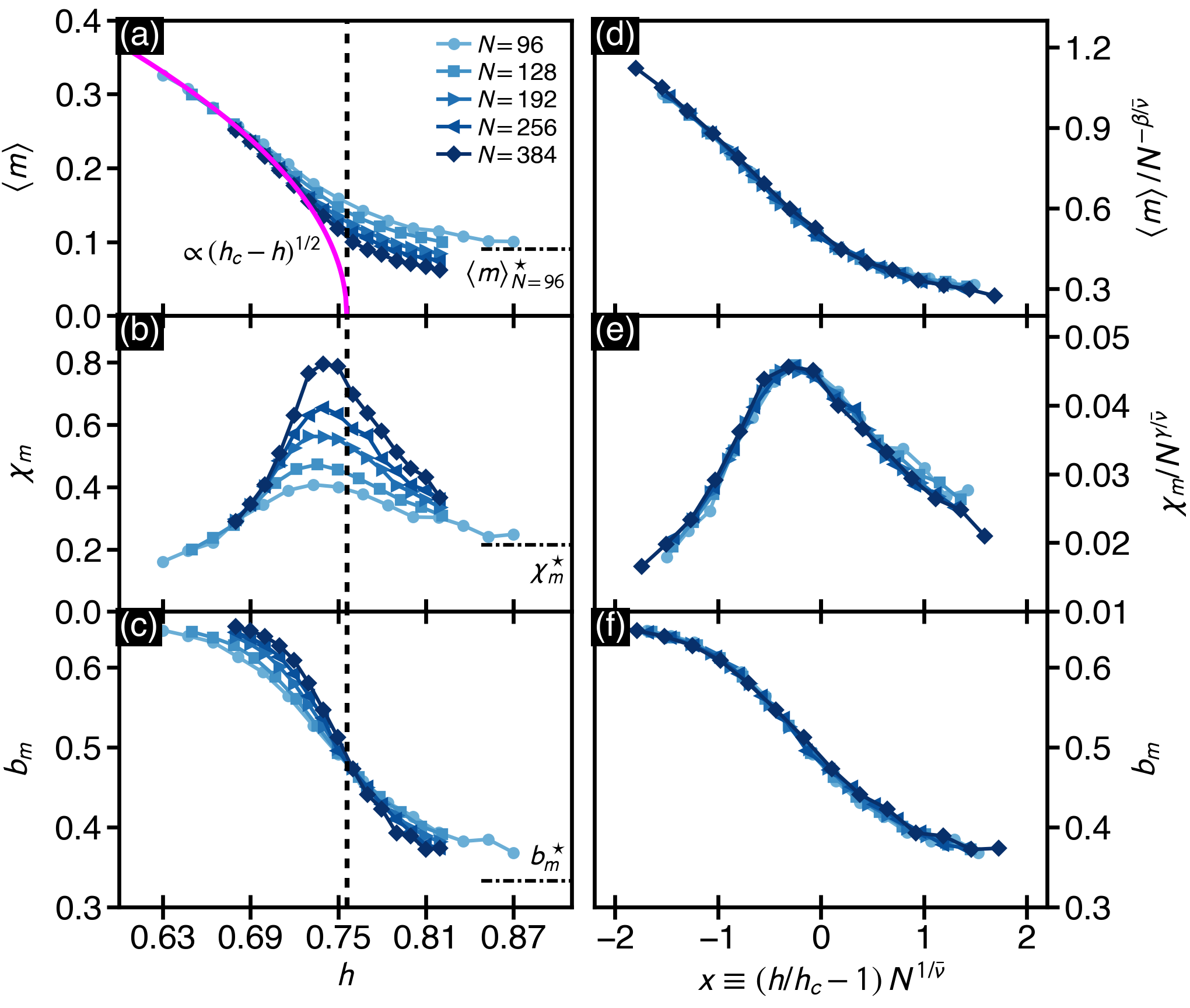}
\caption{
Results of the finite-size scaling analysis at $J_0=1.2$ and $\chi=1$.
(a-c) Scaling behavior as function of energy density $h$. 
(a) Order parameter $\langle m \rangle$. 
Solid line indicates $\langle m \rangle \sim (h_c-h)^{1/2}$ for $h_c=0.756$.
(b) Finite-size susceptibility $\chi_m$.
(c) Binder cumulant $b_m$.
(d-f) Data collapse of scaled quantities
as function of $x\equiv (h/h_c-1)N^{1/\bar{\nu}}$ (for the definition of the parameters see text).
(d) Scaled order parameter.
(e) Scaled susceptibility.
(f) Binder cumulant.
In (a-c), dashed lines indicate $h=0.756$, and dash-dotted lines indicate values expected for $h\to h_{\mathrm{max}}$.
Errorbars are smaller than symbol sizes.
}
\label{fig:3}
\end{figure}

To locate $h_c$, and simultaneously determine $\bar{\nu}$, the dimensionless Binder parameter $b_m$ [Eq.~(\ref{eq:b_m})] is considered first. 
%
At the critical point, according to Eq.~(\ref{eq:scaling_b}), the Binder parameter for the equilibrium system should be independent of $N$ \cite{Binder:ZPB:1981}. 
Figure~\ref{fig:3}(c) shows that indeed, as function of $h$, curves at different $N$ intersect at a common point. 
%
In the limit $x\to \infty$, the numerical results are in reasonable agreement with $b_m^\star=1 - \mu_4^\prime/[3 (\mu_2^\prime)^2] = \frac{1}{3}$ [dash-dotted line in Fig.~\ref{fig:3}(c)], expected on basis of the exact results derived in Sec.~\ref{sec:res_order}.
Let us note that this is different from the corresponding high-temperature limit for the fully connected Ising model, where $b_m \approx 0$ \cite{Challa:PRB:1986,Romano:PRE:2014}.   
%
%
Imposing the scaling form Eq.~(\ref{eq:scaling_b}), and minimizing an objective function measuring the quality of the \emph{data collapse}  \cite{Houdayer:PRB:2004,Houdayer:PRB:2004:S,autoScale:GH:2022}, yields the optimal parameters $h_c=0.756(1)$ and $\bar{\nu}=2.04(8)$.
(An independent analysis, performed by minimizing a local linearity function for Eq.~(\ref{eq:scaling_b}) \cite{Kawashima:JPSJ:1993}, results in the supporting estimates $h_c=0.756$ and $\bar{\nu}=1.97$.)
The excellent data collapse achieved by these parameter values is demonstrated in Fig.~\ref{fig:3}(f).

\subsubsection{\label{sec:res_fss_m}Order parameter}
To locate $h_c$ and simultaneously determine the critical exponents $\beta$ and $\bar{\nu}$, the finite-size scaling of the order parameter is considered next.
%
In the ordered phase close to the critical point, it is well described by $\langle m\rangle \propto |h-h_c|^\beta$.
This is shown by the dashed line in Fig.~\ref{fig:3}(a), obtained for \emph{a priori} choice $h_c=0.756$, and $\beta=1/2$, in qualitative agreement with previous results \cite{Ramos:PRX:2020}.
%
In the limit $x\to \infty$, our numerical results are in excellent agreement with the exact result $\langle m\rangle^\star_N = \sqrt{\pi/4/N}$ obtained in Sec.~\ref{sec:res_order} [dash-dotted line in Fig.~\ref{fig:3}(a)]. 
%
By imposing Eq.~(\ref{eq:scaling_m}), the optimal data collapse quality is achieved for $h_c=0.755(1)$, $\bar{\nu}=2.00(4)$, and $\beta=0.52(4)$
\cite{Houdayer:PRB:2004,Houdayer:PRB:2004:S,autoScale:GH:2022}. 
(Minimizing the local linearity function for Eq.~(\ref{eq:scaling_m}) \cite{Kawashima:JPSJ:1993}, gives $h_c=0.752$, $\bar{\nu}=2.03$, and $\beta=0.51$.) 
%
%
%
The corresponding  data collapse is shown in Fig.~\ref{fig:3}(d).

\subsubsection{\label{sec:res_fss_chi}Finite-size susceptibility}
The finite-size susceptibility $\chi_m$ [Eq.~(\ref{eq:chi_m})] allows to define a sequence of pseudocritical 
points $h_c^\prime$. For a given system size $N$, $h_c^\prime$ is determined by the peak-location of $\chi_m$, see Fig.~\ref{fig:3}(b). At $N=64$ the peak is located at $h_c^\prime = 0.736(2)$. For increasing system size these pseudocritical points shift to larger values of $h$, e.g., at $N=384$ it is $h_c^\prime = 0.744(4)$, approaching the previous estimates of $h_c$ from below. 
%
In the limit $x\to\infty$, the exact results of Sec.~\ref{sec:res_order} lead to expect $\chi_m^\star=N[\mu_2^\prime - (\mu_1^\prime)^2] = 1-\pi/4\approx 0.21$ [dash-dotted line in Fig.~\ref{fig:3}(b)], in good agreement with our numerical results.
%
Imposing Eq.~(\ref{eq:scaling_chi}), the optimal data collapse is achieved
for $h_c=0.750(2)$, $\bar{\nu}=2.00(4)$, and $\gamma=0.98(6)$ \cite{Houdayer:PRB:2004,Houdayer:PRB:2004:S,autoScale:GH:2022}, see Fig.~\ref{fig:3}(e).
(Minimizing the local linearity function for Eq.~(\ref{eq:scaling_chi}) \cite{Kawashima:JPSJ:1993}, yields $h_c=0.753$, $\bar{\nu}=2.06$, and $\gamma=0.99$.)
%
%
%
%
Thus, approaching the critical point from above, the susceptibility obeys the Curie-Weiss law $\chi_m \propto |h-h_c|^{-1}$.



\begin{figure}[tb!]
\centering\includegraphics[width=\linewidth]{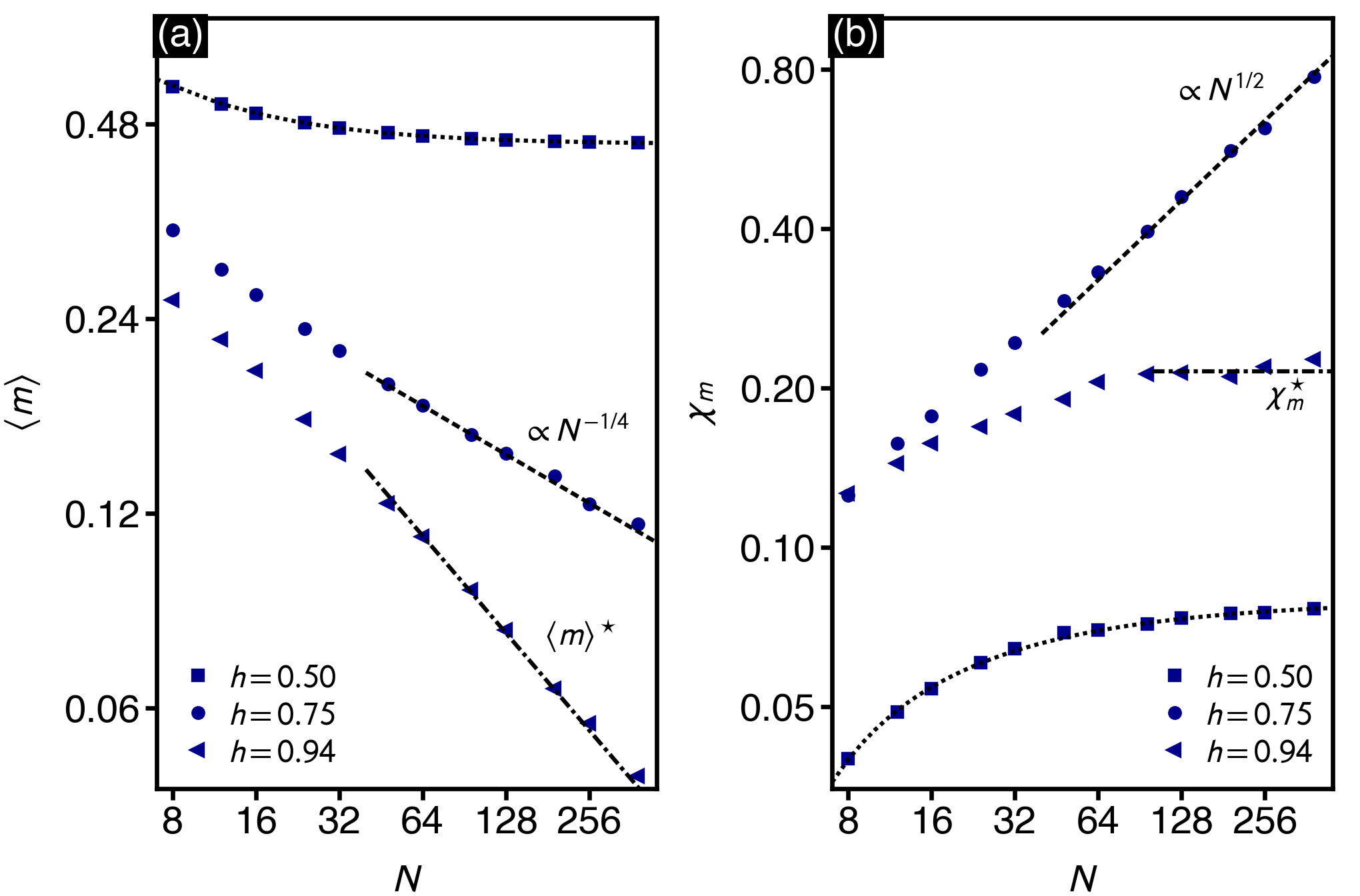}
\caption{Scaling behavior at selected values of $h$.
(a) Order parameter $\langle m \rangle$ in the ordered phase ($h=0.5$), at criticality ($h=0.75\approx h_c$), and in the disordered phase ($h=0.94$).
(b) Same for the finite-size susceptibility $\chi_m$.
Short-dashed lines at $h=0.5$ indicate scaling laws accounting for corrections to scaling (see text).
Dashed lines at $h=0.75$ indicate asymptotic power-law scaling.
Dashed dotted lines at $h=0.94$ indicate exact results. Errorbars are smaller than symbol size.
}
\label{fig:4}
\end{figure}

\begin{figure}[b!]
\centering\includegraphics[width=\linewidth]{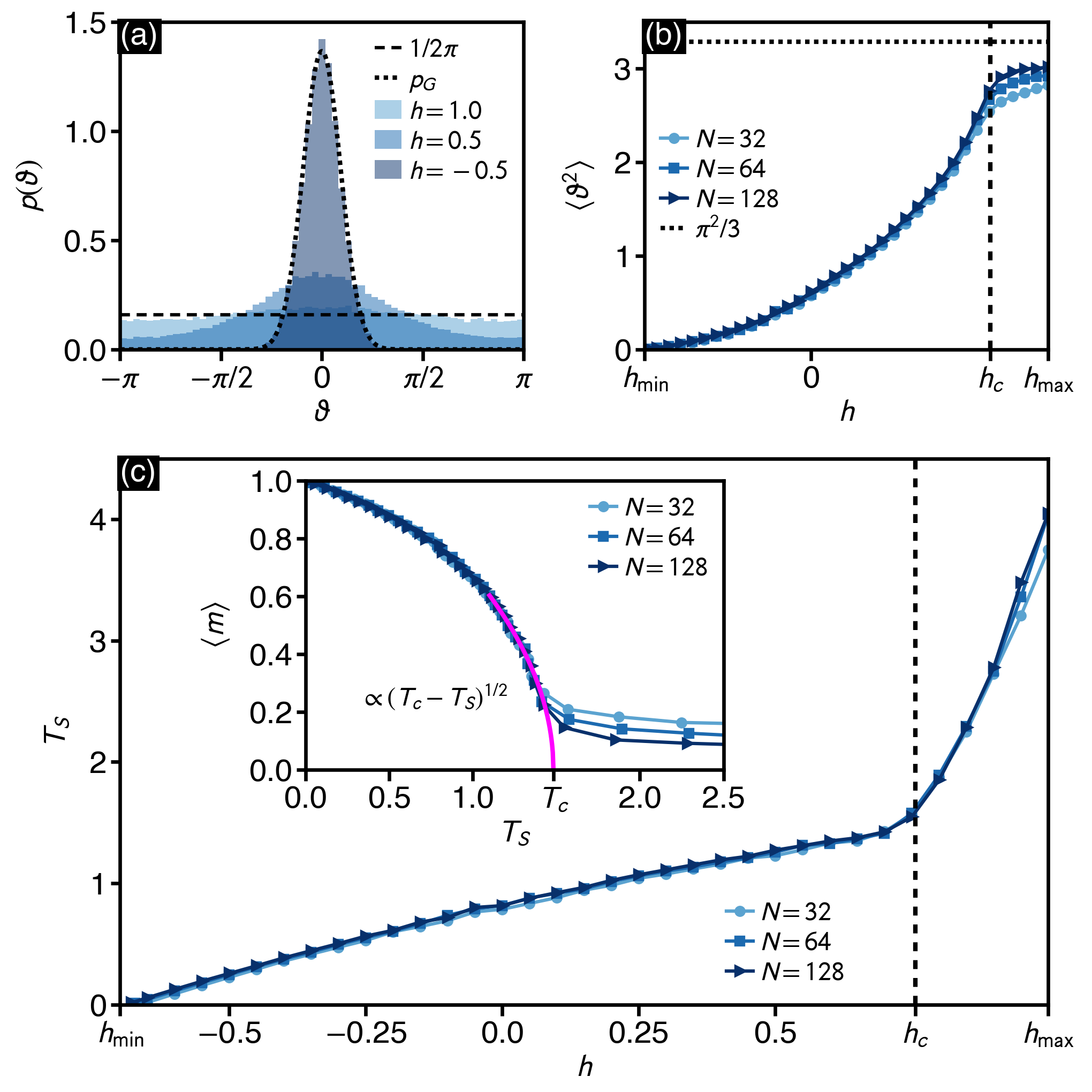}
\caption{Angular displacement and soft-spin temperature.
(a) Probability density function $p(\vartheta)$ of the angular displacement $\vartheta$ of the soft-spins relative to the average spin direction. $p_G$ is a Gaussian with zero mean and variance $\approx 0.085$.
(b) Mean-square angular displacement $\langle \vartheta^2\rangle$ as function of energy density $h$. Vertical dashed line indicates transition point $h_c=0.754$. $\pi^2/3$ is the variance of a uniform distribution in range $(-\pi,\pi)$.
(c) Soft-spin temperature $T_S$ as function of energy density. Inset shows the order parameter as function of soft-spin temperature. The transition occurs at $T_c\approx 1.48$.
}
\label{fig:6}
\end{figure}

\subsection{Scaling behavior at selected energy densities\label{sec:res_scaling_hc}}

In the ordered phase ($h<h_c$) the spontaneous magnetization exhibits the usual scaling behavior of the form $\langle m \rangle = m_\infty(h) + c N^{-x}$ \cite{Botet:PRB:1983}. A least-squares fit at $h=0.5$ for $N>20$ yields $m_\infty(0.5)=0.446(1)$, $c=0.77(6)$, and $x=0.97(3)$, see Fig.~\ref{fig:4}(a).
%
Close to the asymptotic critical point ($h\approx h_c$), power-law scaling $\langle m \rangle = c N^{-x}$,
with $c=0.50(1)$, and $x=0.248(5)$ is observed to hold for $N>30$.
This is consistent with the FSS assumption
[Eq.~(\ref{eq:scaling_m}) for $x\to 0$], supporting an asymptotic exponent ratio $\beta/\bar{\nu}=x=1/4$ [Fig.~\ref{fig:4}(a)]. 
%
Deep in the disordered phase ($h>h_c$), the time-averaged order parameter is expected to scale as $\langle m\rangle^\star = m_0/\sqrt{N}$, where $m_0=\sqrt{\pi/4} \approx 0.89$. Figure~\ref{fig:4}(a) verifies this scaling law for the parameter choice $h=0.94$.
A least squares fit yields $m_{0,{\mathrm{fit}}}=0.88(1)$, in excellent agreement with the expected scaling behavior. 
Reference~\cite{Ramos:PRX:2020} reports the fit results $\bar{m}_{\rm{min}}\approx 0.75/\sqrt{N}$, where, at a given system size $N$, ${\bar{m}}_{\mathrm{min}}$ is the asymptotic value to which $\langle m \rangle$ converges in the disordered phase.
A reason for the smaller scaling-factor might be found by observing that the analysis in Ref.~\cite{Ramos:PRX:2020} is based mostly on systems of size $N<100$, where corrections to scaling might still be large.
For comparison, in the presented study, restricting the fit at $h=0.94$ to systems of size $N<100$ yields $m_{0,{\mathrm{fit}}} = 0.81(2)$, indeed trending towards a smaller value.
%

Similar scaling laws govern the behavior of the finite-size susceptibility $\chi_m$, see Fig.~\ref{fig:4}(b). In the ordered phase at $h=0.5$, the scaling law 
$\chi_m = \chi_{m,\infty}(h) + cN^{-x}$ yields $\chi_{m,\infty}(0.5) = 0.080(1)$, $c=-0.16(1)$, and $x=0.66(2)$.
At the critical point, $\chi_m=c (N+\Delta N)^{-x}$, where $c=0.042(3)$, $\Delta N=5(2)$, and, $x=0.49(1)$, in support of an asymptotic exponent ratio $\gamma/\bar{\nu}=x=0.5$ [cf.~Eq.~\ref{eq:scaling_chi} for $x\to 0$].
In the disordered phase, saturation at $\chi_m^\star = 1-\pi/4\approx 0.21$ is clearly evident for $N>100$.

\subsection{\label{sec:res_temp}Soft-spin temperature}

While in the limit of weak nonlinearity, the framework of optical thermodynamics can be employed to relate the equilibrium properties to a thermodynamic temperature \cite{Ramos:PRX:2020,Wu:NP:2019,Wu:CP:2020,Parto:OL:2019}, no obvious corresponding relation is available in the present case. 
There exist, however, various temperature estimators for use under equilibrium conditions in computer simulations.
For example, a kinetic temperature is defined in molecular dynamics simulations \cite{Allen:BOOK:217}, 
kinetic and spin temperatures are considered for ensembles of spins in semiclassical Langevin dynamics \cite{Ma:PRE:2010,Ma:PRB:2011},
and a configurational temperature is used as diagnostic tool in Monte Carlo simulations \cite{Butler:JCP:1998}. In the microcanonical ensemble, it is further possible to measure temperature in terms of a dynamical, geometrical approach \cite{Rugh:PRL:1997,Rugh:JPA:1998,Nurdin:PA:2002}.

\begin{table}[b!]
\caption{\label{tab:1}%
Critical points and exponents of the optical phase transition on fully connected, equal-coupling photonic networks. From left to right: mode-mode coupling strength ($J_0$), nonlinearity parameter ($\chi$), critical point ($h_c$), coherence number exponent ($\bar{\nu}$), order parameter exponent ($\beta$), and, susceptibility exponent ($\gamma$). Results in the first row are reproduced after Ref.~\cite{Ramos:PRX:2020}.}
\begin{ruledtabular}
\begin{tabular}{cccccc}
$J_0$ & $\chi$ & $h_c$      & $\bar{\nu}$   & $\beta$   & $\gamma$  \\
\hline
1.2   & 1.0    & 0.75       & --            & $\frac{1}{2}$       &  --       \\
1.2   & 0.6    & 0.47(1)    & 2.0(1)        & 0.49(1)   & 0.92(8)   \\
1.2   & 0.8    & 0.619(5)   & 2.00(4)       & 0.52(1)   & 0.95(5)   \\
1.2   & 1.0    & 0.754(4)   & 2.00(4)       & 0.52(4)   & 0.98(6)   \\
1.2   & 1.2    & 0.880(1)   & 2.00(2)       & 0.50(1)   & 0.99(2)   \\
\end{tabular}
\end{ruledtabular}
\end{table}

Here we introduce an expression for the temperature, based entirely on the angular momenta $p_\ell \equiv |{\mathbf{s}}_\ell| \,\dot{\vartheta}_\ell$ of the photonic soft-spins, where $\vartheta_\ell\equiv \theta_\ell - \theta$ measures the angular displacement of soft-spin ${\mathbf{s}_\ell}$ relative to ${\mathbf{m}}$.
The corresponding soft-spin temperature $T_S$ is computed as the time-averaged, squared angular momentum per spin
\begin{align}
T_S = \frac{1}{N} \sum_\ell \langle p_\ell^2 \rangle,\label{eq:Ts}
\end{align}
%
with the system in equilibrium.
Deep in the ordered phase, the distribution of the angles $\vartheta_\ell$ is a narrow Gaussian with zero mean, e.g., at $h=-0.5$ its variance is $\langle \vartheta^2\rangle\approx 0.085$, see Fig.~\ref{fig:6}(a). For increasing energy density, the variance increases. In the disordered phase, close to $h_{\rm{max}}$, it compares well to a uniform distribution $p_{\rm{u}}(\vartheta)=1/(2\pi)$ in range $-\pi \leq \vartheta \leq \pi$. The increase of $\langle \vartheta^2\rangle$ with $h$ is shown in Fig.~\ref{fig:6}(b).
In the disordered phase, the variance $\int_{-\pi}^{\pi} \vartheta^2\, p_{\rm{u}}(\vartheta)~{\mathrm{d}}\vartheta=\pi^2/3$ is approached as $N\to \infty$. The inflection points of the curves are located close to the transition point $h_c$.
It is not surprising that the angular displacements $\vartheta_\ell$ are a sensitive measure of orientational order in the system: in case of the XY model, at low temperatures, a direct relation between a quantity similar to $1-\vartheta^2$ and the magnetization can be established \cite{Tobochnik:PRB:1979,Ota:JPC:1992}. 
Similar to the temperature considered in earlier studies of the XY model \cite{Leoncini:PRE:1998}, $T_S$ is associated with kinetic information contained in the rotational degrees of freedom of the system.
%
%
Figure~\ref{fig:6}(c) indicates a monotonous increase of the soft-spin temperature with energy density. In the limit $h\to h_{\rm{min}}$, $T_S\to 0$.
Further, for decreasing soft-spin temperature, the length fluctuations of the spins are increasingly suppressed, strengthening the analogy to the XY model.
%
%
Considering $T_S$ as control parameter, we find that in the vicinity of the critical point 
$\langle m \rangle \propto |T_S-T_c|^{1/2}$, see inset of Fig.~\ref{fig:6}(c), according to which the order parameter vanishes with exponent $\beta=1/2$, in agreement with the mean-field theory for photonic networks developed in Ref.~\cite{Ramos:PRX:2020}.
This is explained by $T_S$ being a linear function of $h$ for $h\lessapprox h_c$.
Hence, a consistent description of the phase transition is also obtained in terms of the soft-spin temperature.
%

\begin{figure}[tb!]
\centering\includegraphics[width=\linewidth]{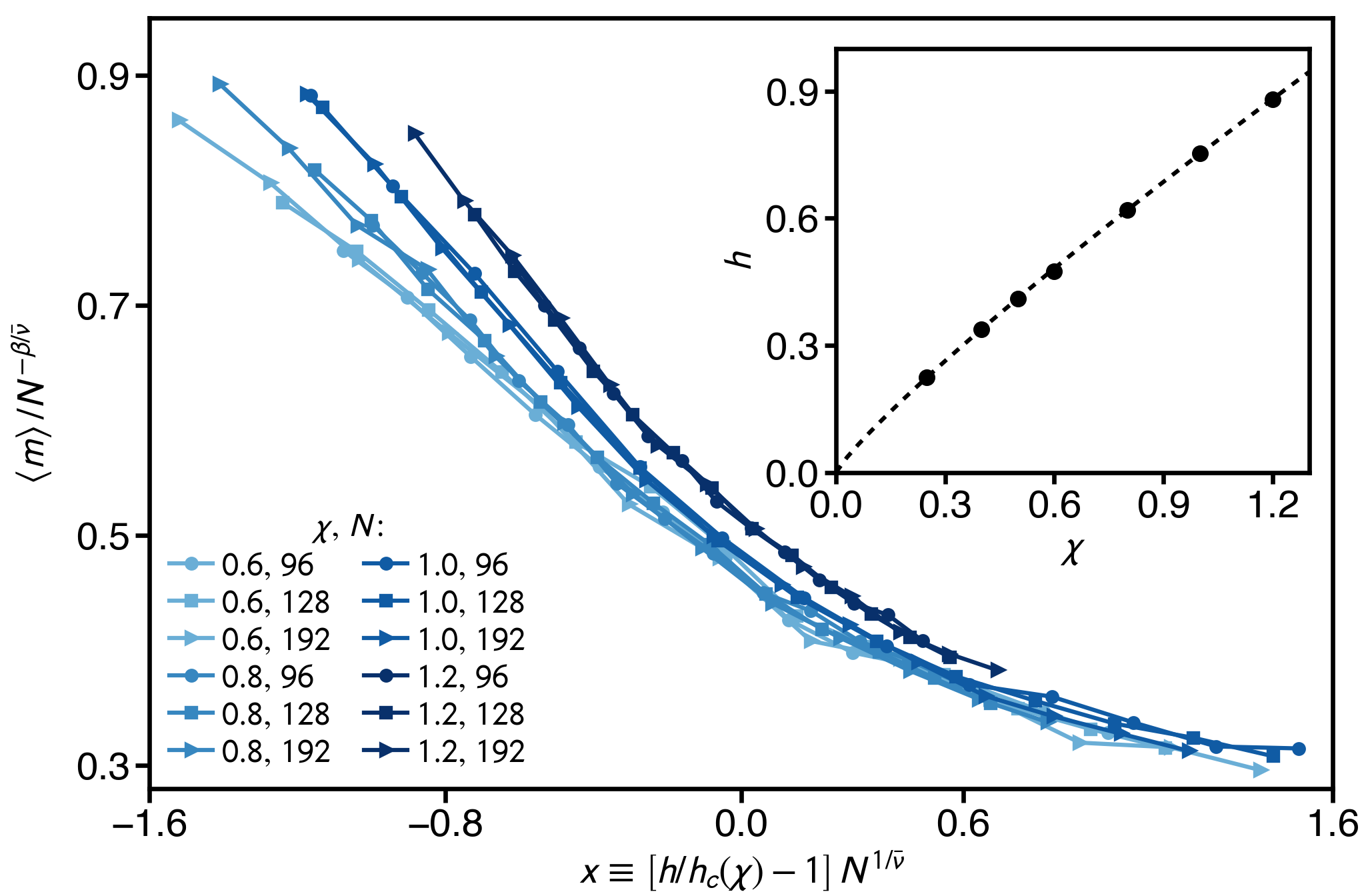}
\caption{Finite-size scaling results at different values of the nonlinearity parameter $\chi$.
The main figure shows the rescaled magnetization as function of the scaling variable $x\equiv \left[h/h_c(\chi) -1\right] N^{1/\bar{\nu}}$.
The inset shows a phase diagram in the $\chi$-$h$ plane, indicating
the $\chi$-dependence $h_c(\chi) = 0.754 \,\chi^{0.88}$ of the transition point.
}
\label{fig:5}
\end{figure}

\section{\label{sec:d_and_c} Discussion and conclusions}

In the present study, we build upon a recent work \cite{Ramos:PRX:2020}, wherein a magnetization-like order parameter $m$ was considered for studying optical phase transitions in photonic networks.
Going beyond previous results, we showed that the two-component quantity $\mathbf{m}$, a two-component order parameter inspired by classical statistical mechanics models, provides a deeper insight into the equilibrium dynamics. 
Within the ordered phase it has the additional benefit of distinguishing between non-identical equilibrium configurations at fixed energy density, revealing collective angular excitations sustained by the photonic soft-spins. These are a manifestation of a symmetry of the Hamiltonian in the system dynamics. 
%
Within the disordered phase, it even allows to obtain exact results for several quantities of interest. Specifically, it predicts the order parameter to follow a Rayleigh-distribution, and even yields an expression for the finite-size scaling behavior of $\langle m\rangle$ in the limit $h\to h_{\mathrm{max}}$, both in excellent agreement with our numerical simulations.
Qualitatively we find that the large-$N$ behavior of the order parameter in equal-coupling photonic networks is consistent with that of statistics mechanics mean-field spin systems, such as the fully connected Ising model \cite{Botet:PRB:1983}, and the XY model on small-world networks \cite{Kim:PRE:2001}, and long-range interacting one-dimensional systems \cite{Hong:PRE:2015}.

%
Let us note that, in the vicinity of the asymptotic critical point, the order parameter distribution is not Gaussian and has a maximum at a nonzero value of $m$, see Figs.~\ref{fig:1}(b,d) and the Supplementary Material \cite{SuppMat}.
%
In case of the fully connected Ising model this has led to the observation of anomalous mean-field scaling \cite{Romano:PRE:2014}, where, at criticality, the mean value of the magnetization and the most probable magnetization were found to scale differently.
%
In the present case we verified that the scaling of the mean value and the most probable value of $m$ are identical (not shown). 
%

%
In addition to the FSS analysis reported in Sec.~\ref{sec:res_fss}, we performed further, similar analysis for other choices of the nonlinearity parameter $\chi$. 
To account for these results, Fig.~\ref{fig:5} shows the data collapse achieved for the order parameter, when taking into account a $\chi$-dependent critical point $h_c(\chi)$.
In either case, the critical exponents where found to be in good agreement with the above analysis, see Tab.~\ref{tab:1}.
We find that the location of the critical point is well represented by the critical line $h_c(\chi) = 0.754 \cdot \chi^x$, with sub-linear exponent $x\approx 0.88$, see the $\chi$-$h$ phase diagram shown in the inset of Fig.~\ref{fig:5}. 
%

%


%

%

%

%
%
In conclusion, we have studied fully connected ECPNs using dynamic simulations of the nonlinear equations of motion of the coupled modes. 
For several choices of the nonlinearity parameter $\chi$, critical points and exponents have been determined by imposing appropriate scaling assumptions on a magnetization-like order parameter, its associated Binder cumulant and finite-size susceptibility.
For all parameter settings considered in this work, the FSS analysis of the optical phase transition yields mean-field critical exponents and allows to infer the upper critical dimension $d_c=4$.
Overall, corrections to scaling where found to affect the scaling behavior for systems of size up to $N\approx 100$.
Generally, for systems of finite size, finite-size effects result in effective transition points that differ slightly from the asymptotic values quoted in Tab.~\ref{tab:1}, also see the discussion in Sec.~\ref{sec:res_fss_chi}.
These findings may prove to be useful, when considering instances of photonic networks with finite (small) number of building blocks in optical technologies.
We further established a connection between the energy density and a kinetic temperature, measuring the activation of the angular degree of freedom of the photonic soft spins, in terms of which a consistent signature of the transition could be obtained.
Finally, let us note that in Ref.~\cite{Ramos:PRX:2020}, a mean-field theory for photonic networks with finite coordination number was formulated. 
From the point of view of statistical mechanics it would be rewarding to test this mean-field theory by studying the equal-coupling photonic network on random regular graphs \cite{Bollobas:BOOK:2001,Melchert:PRE:2011}, an ensemble of random graphs with fixed coordination number. 

\begin{acknowledgments}
I would like to thank Ayhan Demircan for helpful discussions.
Funding by the Deutsche Forschungsgemeinschaft within the Cluster of Excellence
PhoenixD (Photonics, Optics, and Engineering–Innovation Across Disciplines)
(EXC 2122, projectID 390833453) is gratefully acknowledged.
\end{acknowledgments}

\bibliography{references}

\begin{thebibliography}{51}%
\makeatletter
\providecommand \@ifxundefined [1]{%
 \@ifx{#1\undefined}
}%
\providecommand \@ifnum [1]{%
 \ifnum #1\expandafter \@firstoftwo
 \else \expandafter \@secondoftwo
 \fi
}%
\providecommand \@ifx [1]{%
 \ifx #1\expandafter \@firstoftwo
 \else \expandafter \@secondoftwo
 \fi
}%
\providecommand \natexlab [1]{#1}%
\providecommand \enquote  [1]{``#1''}%
\providecommand \bibnamefont  [1]{#1}%
\providecommand \bibfnamefont [1]{#1}%
\providecommand \citenamefont [1]{#1}%
\providecommand \href@noop [0]{\@secondoftwo}%
\providecommand \href [0]{\begingroup \@sanitize@url \@href}%
\providecommand \@href[1]{\@@startlink{#1}\@@href}%
\providecommand \@@href[1]{\endgroup#1\@@endlink}%
\providecommand \@sanitize@url [0]{\catcode `\\12\catcode `\$12\catcode
  `\&12\catcode `\#12\catcode `\^12\catcode `\_12\catcode `\%12\relax}%
\providecommand \@@startlink[1]{}%
\providecommand \@@endlink[0]{}%
\providecommand \url  [0]{\begingroup\@sanitize@url \@url }%
\providecommand \@url [1]{\endgroup\@href {#1}{\urlprefix }}%
\providecommand \urlprefix  [0]{URL }%
\providecommand \Eprint [0]{\href }%
\providecommand \doibase [0]{https://doi.org/}%
\providecommand \selectlanguage [0]{\@gobble}%
\providecommand \bibinfo  [0]{\@secondoftwo}%
\providecommand \bibfield  [0]{\@secondoftwo}%
\providecommand \translation [1]{[#1]}%
\providecommand \BibitemOpen [0]{}%
\providecommand \bibitemStop [0]{}%
\providecommand \bibitemNoStop [0]{.\EOS\space}%
\providecommand \EOS [0]{\spacefactor3000\relax}%
\providecommand \BibitemShut  [1]{\csname bibitem#1\endcsname}%
\let\auto@bib@innerbib\@empty
\bibitem [{\citenamefont {Cardy}(1996)}]{Cardy:BOOK:1996}%
  \BibitemOpen
  \bibfield  {author} {\bibinfo {author} {\bibfnamefont {J.}~\bibnamefont
  {Cardy}},\ }\href@noop {} {\emph {\bibinfo {title} {{Scaling and
  Renormalization in Statistical Physics}}}}\ (\bibinfo  {publisher} {Cambridge
  University Press},\ \bibinfo {year} {1996})\BibitemShut {NoStop}%
\bibitem [{\citenamefont {Newman}\ and\ \citenamefont
  {Barkema}(1999)}]{Newman:BOOK:1999}%
  \BibitemOpen
  \bibfield  {author} {\bibinfo {author} {\bibfnamefont {M.~E.~J.}\
  \bibnamefont {Newman}}\ and\ \bibinfo {author} {\bibfnamefont {G.~T.}\
  \bibnamefont {Barkema}},\ }\href@noop {} {\emph {\bibinfo {title} {{Monte
  Carlo Methods in Statistical Physics}}}}\ (\bibinfo  {publisher} {Oxford
  University Press},\ \bibinfo {year} {1999})\BibitemShut {NoStop}%
\bibitem [{\citenamefont {Binder}\ and\ \citenamefont
  {Heermann}(2010)}]{Binder:BOOK:2010}%
  \BibitemOpen
  \bibfield  {author} {\bibinfo {author} {\bibfnamefont {K.}~\bibnamefont
  {Binder}}\ and\ \bibinfo {author} {\bibfnamefont {D.~W.}\ \bibnamefont
  {Heermann}},\ }\href@noop {} {\emph {\bibinfo {title} {{Monte Carlo
  Simulation in Statistical Physics}}}}\ (\bibinfo  {publisher} {Springer},\
  \bibinfo {year} {2010})\BibitemShut {NoStop}%
\bibitem [{\citenamefont {Hartmann}(2001)}]{Hartmann:BOOK:2001}%
  \BibitemOpen
  \bibfield  {author} {\bibinfo {author} {\bibfnamefont {A.~K.}\ \bibnamefont
  {Hartmann}},\ }\href@noop {} {\emph {\bibinfo {title} {{Optimization
  Algorithms in Physics}}}}\ (\bibinfo  {publisher} {Wiley-VCH},\ \bibinfo
  {year} {2001})\BibitemShut {NoStop}%
\bibitem [{\citenamefont {Binder}(1981)}]{Binder:ZPB:1981}%
  \BibitemOpen
  \bibfield  {author} {\bibinfo {author} {\bibfnamefont {K.}~\bibnamefont
  {Binder}},\ }\bibfield  {title} {\bibinfo {title} {{Finite size scaling
  analysis of ising model block distribution functions}},\ }\href@noop {}
  {\bibfield  {journal} {\bibinfo  {journal} {Z. Phys. B Condens. Matter}\
  }\textbf {\bibinfo {volume} {43}},\ \bibinfo {pages} {119} (\bibinfo {year}
  {1981})}\BibitemShut {NoStop}%
\bibitem [{\citenamefont {Stanley}(1999)}]{Stanley:RMP:1999}%
  \BibitemOpen
  \bibfield  {author} {\bibinfo {author} {\bibfnamefont {H.~E.}\ \bibnamefont
  {Stanley}},\ }\bibfield  {title} {\bibinfo {title} {{Scaling, universality,
  and renormalization: Three pillars of modern critical phenomena}},\
  }\href@noop {} {\bibfield  {journal} {\bibinfo  {journal} {Rev. Mod. Phys.}\
  }\textbf {\bibinfo {volume} {71}},\ \bibinfo {pages} {358} (\bibinfo {year}
  {1999})}\BibitemShut {NoStop}%
\bibitem [{\citenamefont {Stanley}(1987)}]{Stanley:BOOK:1987}%
  \BibitemOpen
  \bibfield  {author} {\bibinfo {author} {\bibfnamefont {H.~E.}\ \bibnamefont
  {Stanley}},\ }\href@noop {} {\emph {\bibinfo {title} {{Introduction to Phase
  Transitions and Critical Phenomena}}}}\ (\bibinfo  {publisher} {Oxford
  University Press},\ \bibinfo {year} {1987})\BibitemShut {NoStop}%
\bibitem [{\citenamefont {Ramos}\ \emph {et~al.}(2020)\citenamefont {Ramos},
  \citenamefont {Fern\'andez-Alc\'azar}, \citenamefont {Kottos},\ and\
  \citenamefont {Shapiro}}]{Ramos:PRX:2020}%
  \BibitemOpen
  \bibfield  {author} {\bibinfo {author} {\bibfnamefont {A.}~\bibnamefont
  {Ramos}}, \bibinfo {author} {\bibfnamefont {L.}~\bibnamefont
  {Fern\'andez-Alc\'azar}}, \bibinfo {author} {\bibfnamefont {T.}~\bibnamefont
  {Kottos}},\ and\ \bibinfo {author} {\bibfnamefont {B.}~\bibnamefont
  {Shapiro}},\ }\bibfield  {title} {\bibinfo {title} {{Optical Phase
  Transitions in Photonic Networks: a Spin-System Formulation}},\ }\href@noop
  {} {\bibfield  {journal} {\bibinfo  {journal} {Phys. Rev. X}\ }\textbf
  {\bibinfo {volume} {10}},\ \bibinfo {pages} {031024} (\bibinfo {year}
  {2020})}\BibitemShut {NoStop}%
\bibitem [{\citenamefont {Stanley}(1968)}]{Stanley:PRL:1968}%
  \BibitemOpen
  \bibfield  {author} {\bibinfo {author} {\bibfnamefont {H.~E.}\ \bibnamefont
  {Stanley}},\ }\bibfield  {title} {\bibinfo {title} {{Dependence of Critical
  Properties on Dimensionality of Spins}},\ }\href@noop {} {\bibfield
  {journal} {\bibinfo  {journal} {Phys. Rev. Lett.}\ }\textbf {\bibinfo
  {volume} {20}},\ \bibinfo {pages} {589} (\bibinfo {year} {1968})}\BibitemShut
  {NoStop}%
\bibitem [{\citenamefont {Tobochnik}\ and\ \citenamefont
  {Chester}(1979)}]{Tobochnik:PRB:1979}%
  \BibitemOpen
  \bibfield  {author} {\bibinfo {author} {\bibfnamefont {J.}~\bibnamefont
  {Tobochnik}}\ and\ \bibinfo {author} {\bibfnamefont {G.~V.}\ \bibnamefont
  {Chester}},\ }\bibfield  {title} {\bibinfo {title} {{Monte Carlo study of the
  planar spin model}},\ }\href@noop {} {\bibfield  {journal} {\bibinfo
  {journal} {Phys. Rev. B}\ }\textbf {\bibinfo {volume} {20}},\ \bibinfo
  {pages} {3761} (\bibinfo {year} {1979})}\BibitemShut {NoStop}%
\bibitem [{XY:()}]{XY:NOTE}%
  \BibitemOpen
  \href@noop {} {}\bibinfo {note} {In the XY model, spins are two-component
  vectors restricted to unit length, able to point in any direction on a
  two-dimensional plane \cite{Cardy:BOOK:1996}.}\BibitemShut {Stop}%
\bibitem [{\citenamefont {Kim}\ \emph {et~al.}(2001)\citenamefont {Kim},
  \citenamefont {Hong}, \citenamefont {Holme}, \citenamefont {Jeon},
  \citenamefont {Minnhagen},\ and\ \citenamefont {Choi}}]{Kim:PRE:2001}%
  \BibitemOpen
  \bibfield  {author} {\bibinfo {author} {\bibfnamefont {B.~J.}\ \bibnamefont
  {Kim}}, \bibinfo {author} {\bibfnamefont {H.}~\bibnamefont {Hong}}, \bibinfo
  {author} {\bibfnamefont {P.}~\bibnamefont {Holme}}, \bibinfo {author}
  {\bibfnamefont {G.~S.}\ \bibnamefont {Jeon}}, \bibinfo {author}
  {\bibfnamefont {P.}~\bibnamefont {Minnhagen}},\ and\ \bibinfo {author}
  {\bibfnamefont {M.~Y.}\ \bibnamefont {Choi}},\ }\bibfield  {title} {\bibinfo
  {title} {{XY model in small-world networks}},\ }\href@noop {} {\bibfield
  {journal} {\bibinfo  {journal} {Phys. Rev. E}\ }\textbf {\bibinfo {volume}
  {64}},\ \bibinfo {pages} {056135} (\bibinfo {year} {2001})}\BibitemShut
  {NoStop}%
\bibitem [{\citenamefont {Hong}\ and\ \citenamefont
  {Kim}(2015)}]{Hong:PRE:2015}%
  \BibitemOpen
  \bibfield  {author} {\bibinfo {author} {\bibfnamefont {H.}~\bibnamefont
  {Hong}}\ and\ \bibinfo {author} {\bibfnamefont {B.~J.}\ \bibnamefont {Kim}},\
  }\bibfield  {title} {\bibinfo {title} {{Winding number excitation detects
  phase transition in one-dimensional $XY$ model with variable interaction
  range}},\ }\href@noop {} {\bibfield  {journal} {\bibinfo  {journal} {Phys.
  Rev. E}\ }\textbf {\bibinfo {volume} {91}},\ \bibinfo {pages} {052120}
  (\bibinfo {year} {2015})}\BibitemShut {NoStop}%
\bibitem [{\citenamefont {Small}\ \emph {et~al.}(2011)\citenamefont {Small},
  \citenamefont {Pugatch},\ and\ \citenamefont {Silberberg}}]{Small:PRA:2011}%
  \BibitemOpen
  \bibfield  {author} {\bibinfo {author} {\bibfnamefont {E.}~\bibnamefont
  {Small}}, \bibinfo {author} {\bibfnamefont {R.}~\bibnamefont {Pugatch}},\
  and\ \bibinfo {author} {\bibfnamefont {Y.}~\bibnamefont {Silberberg}},\
  }\bibfield  {title} {\bibinfo {title} {Berezinskii-kosterlitz-thouless
  crossover in a photonic lattice},\ }\href@noop {} {\bibfield  {journal}
  {\bibinfo  {journal} {Phys. Rev. A}\ }\textbf {\bibinfo {volume} {83}},\
  \bibinfo {pages} {013806} (\bibinfo {year} {2011})}\BibitemShut {NoStop}%
\bibitem [{\citenamefont {Shi}\ \emph {et~al.}(2021)\citenamefont {Shi},
  \citenamefont {Kottos},\ and\ \citenamefont {Shapiro}}]{Shi:PRR:2021}%
  \BibitemOpen
  \bibfield  {author} {\bibinfo {author} {\bibfnamefont {C.}~\bibnamefont
  {Shi}}, \bibinfo {author} {\bibfnamefont {T.}~\bibnamefont {Kottos}},\ and\
  \bibinfo {author} {\bibfnamefont {B.}~\bibnamefont {Shapiro}},\ }\bibfield
  {title} {\bibinfo {title} {Controlling optical beam thermalization via
  band-gap engineering},\ }\href@noop {} {\bibfield  {journal} {\bibinfo
  {journal} {Phys. Rev. Research}\ }\textbf {\bibinfo {volume} {3}},\ \bibinfo
  {pages} {033219} (\bibinfo {year} {2021})}\BibitemShut {NoStop}%
\bibitem [{\citenamefont {Wu}\ \emph {et~al.}(2019)\citenamefont {Wu},
  \citenamefont {Hassan},\ and\ \citenamefont {Christodoulides}}]{Wu:NP:2019}%
  \BibitemOpen
  \bibfield  {author} {\bibinfo {author} {\bibfnamefont {F.}~\bibnamefont
  {Wu}}, \bibinfo {author} {\bibfnamefont {A.}~\bibnamefont {Hassan}},\ and\
  \bibinfo {author} {\bibfnamefont {D.}~\bibnamefont {Christodoulides}},\
  }\bibfield  {title} {\bibinfo {title} {Thermodynamic theory of highly
  multimoded nonlinear optical systems},\ }\href@noop {} {\bibfield  {journal}
  {\bibinfo  {journal} {Nat. Photonics}\ }\textbf {\bibinfo {volume} {13}},\
  \bibinfo {pages} {776} (\bibinfo {year} {2019})}\BibitemShut {NoStop}%
\bibitem [{\citenamefont {Wu}\ \emph {et~al.}(2020)\citenamefont {Wu},
  \citenamefont {Jung}, \citenamefont {Parto}, \citenamefont {Khajavikhan},\
  and\ \citenamefont {Christodoulides}}]{Wu:CP:2020}%
  \BibitemOpen
  \bibfield  {author} {\bibinfo {author} {\bibfnamefont {F.}~\bibnamefont
  {Wu}}, \bibinfo {author} {\bibfnamefont {P.}~\bibnamefont {Jung}}, \bibinfo
  {author} {\bibfnamefont {M.}~\bibnamefont {Parto}}, \bibinfo {author}
  {\bibfnamefont {M.}~\bibnamefont {Khajavikhan}},\ and\ \bibinfo {author}
  {\bibfnamefont {D.~N.}\ \bibnamefont {Christodoulides}},\ }\bibfield  {title}
  {\bibinfo {title} {Entropic thermodynamics of nonlinear photonic chain
  networks},\ }\href@noop {} {\bibfield  {journal} {\bibinfo  {journal}
  {Commun. Phys.}\ }\textbf {\bibinfo {volume} {3}},\ \bibinfo {pages} {216}
  (\bibinfo {year} {2020})}\BibitemShut {NoStop}%
\bibitem [{\citenamefont {Parto}\ \emph {et~al.}(2019)\citenamefont {Parto},
  \citenamefont {Wu}, \citenamefont {Jung}, \citenamefont {Makris},\ and\
  \citenamefont {Christodoulides}}]{Parto:OL:2019}%
  \BibitemOpen
  \bibfield  {author} {\bibinfo {author} {\bibfnamefont {M.}~\bibnamefont
  {Parto}}, \bibinfo {author} {\bibfnamefont {F.~O.}\ \bibnamefont {Wu}},
  \bibinfo {author} {\bibfnamefont {P.~S.}\ \bibnamefont {Jung}}, \bibinfo
  {author} {\bibfnamefont {K.}~\bibnamefont {Makris}},\ and\ \bibinfo {author}
  {\bibfnamefont {D.~N.}\ \bibnamefont {Christodoulides}},\ }\bibfield  {title}
  {\bibinfo {title} {Thermodynamic conditions governing the optical temperature
  and chemical potential in nonlinear highly multimoded photonic systems},\
  }\href@noop {} {\bibfield  {journal} {\bibinfo  {journal} {Opt. Lett.}\
  }\textbf {\bibinfo {volume} {44}},\ \bibinfo {pages} {3936} (\bibinfo {year}
  {2019})}\BibitemShut {NoStop}%
\bibitem [{\citenamefont {Yamaguchi}\ \emph {et~al.}(2019)\citenamefont
  {Yamaguchi}, \citenamefont {Das},\ and\ \citenamefont
  {Gupta}}]{Yamaguchi:PRE:2019}%
  \BibitemOpen
  \bibfield  {author} {\bibinfo {author} {\bibfnamefont {Y.~Y.}\ \bibnamefont
  {Yamaguchi}}, \bibinfo {author} {\bibfnamefont {D.}~\bibnamefont {Das}},\
  and\ \bibinfo {author} {\bibfnamefont {S.}~\bibnamefont {Gupta}},\ }\bibfield
   {title} {\bibinfo {title} {Critical exponents in mean-field classical spin
  systems},\ }\href@noop {} {\bibfield  {journal} {\bibinfo  {journal} {Phys.
  Rev. E}\ }\textbf {\bibinfo {volume} {100}},\ \bibinfo {pages} {032131}
  (\bibinfo {year} {2019})}\BibitemShut {NoStop}%
\bibitem [{\citenamefont {Leoncini}\ \emph {et~al.}(1998)\citenamefont
  {Leoncini}, \citenamefont {Verga},\ and\ \citenamefont
  {Ruffo}}]{Leoncini:PRE:1998}%
  \BibitemOpen
  \bibfield  {author} {\bibinfo {author} {\bibfnamefont {X.}~\bibnamefont
  {Leoncini}}, \bibinfo {author} {\bibfnamefont {A.~D.}\ \bibnamefont
  {Verga}},\ and\ \bibinfo {author} {\bibfnamefont {S.}~\bibnamefont {Ruffo}},\
  }\bibfield  {title} {\bibinfo {title} {{Hamiltonian dynamics and the phase
  transition of the $\mathrm{XY}$ model}},\ }\href@noop {} {\bibfield
  {journal} {\bibinfo  {journal} {Phys. Rev. E}\ }\textbf {\bibinfo {volume}
  {57}},\ \bibinfo {pages} {6377} (\bibinfo {year} {1998})}\BibitemShut
  {NoStop}%
\bibitem [{\citenamefont {Hairer}\ \emph {et~al.}(1993)\citenamefont {Hairer},
  \citenamefont {Norsett},\ and\ \citenamefont {Wanner}}]{Hairer:BOOK:1993}%
  \BibitemOpen
  \bibfield  {author} {\bibinfo {author} {\bibfnamefont {E.}~\bibnamefont
  {Hairer}}, \bibinfo {author} {\bibfnamefont {S.~P.}\ \bibnamefont
  {Norsett}},\ and\ \bibinfo {author} {\bibfnamefont {G.}~\bibnamefont
  {Wanner}},\ }\href@noop {} {\emph {\bibinfo {title} {{Solving ordinary
  Differential Equations I. Nonstiff Problems}}}}\ (\bibinfo  {publisher}
  {Springer},\ \bibinfo {year} {1993})\BibitemShut {NoStop}%
\bibitem [{\citenamefont {Melchert}(2022)}]{pyecpn:GH:2022}%
  \BibitemOpen
  \bibfield  {author} {\bibinfo {author} {\bibfnamefont {O.}~\bibnamefont
  {Melchert}},\ }\href@noop {} {\bibinfo {title} {{py-ecpn: Python tools for
  studying mean-field equal-coupling photonic networks}}},\ \bibinfo
  {howpublished} {\url{https://github.com/omelchert/py-ecpn.git}} (\bibinfo
  {year} {2022})\BibitemShut {NoStop}%
\bibitem [{DOP()}]{DOP853:NOTE}%
  \BibitemOpen
  \href@noop {} {}\bibinfo {note} {We use the step size controlled Runge-Kutta
  method of order $8(5,3)$, called {\tt{DOP853}} \cite{Hairer:BOOK:1993},
  wherein the relative single-step accuracy is controlled by a parameter
  $\delta$. For our numerical simulations we use
  $\delta=10^{-11}$.}\BibitemShut {Stop}%
\bibitem [{Sup()}]{SuppMat}%
  \BibitemOpen
  \href@noop {} {}\bibinfo {note} {See Supplemental Material at [URL will be
  inserted by publisher] for details.}\BibitemShut {Stop}%
\bibitem [{\citenamefont {Goldstone}\ \emph {et~al.}(1962)\citenamefont
  {Goldstone}, \citenamefont {Salam},\ and\ \citenamefont
  {Weinberg}}]{Goldstone:PR:1962}%
  \BibitemOpen
  \bibfield  {author} {\bibinfo {author} {\bibfnamefont {J.}~\bibnamefont
  {Goldstone}}, \bibinfo {author} {\bibfnamefont {A.}~\bibnamefont {Salam}},\
  and\ \bibinfo {author} {\bibfnamefont {S.}~\bibnamefont {Weinberg}},\
  }\bibfield  {title} {\bibinfo {title} {{Broken Symmetries}},\ }\href@noop {}
  {\bibfield  {journal} {\bibinfo  {journal} {Phys. Rev.}\ }\textbf {\bibinfo
  {volume} {127}},\ \bibinfo {pages} {965} (\bibinfo {year}
  {1962})}\BibitemShut {NoStop}%
\bibitem [{\citenamefont {Filho}\ and\ \citenamefont
  {Marcos}(2020)}]{Filho:PRE:2020}%
  \BibitemOpen
  \bibfield  {author} {\bibinfo {author} {\bibfnamefont {T.~M.~R.}\
  \bibnamefont {Filho}}\ and\ \bibinfo {author} {\bibfnamefont
  {B.}~\bibnamefont {Marcos}},\ }\bibfield  {title} {\bibinfo {title}
  {{Classical Goldstone modes in long-range interacting systems}},\ }\href@noop
  {} {\bibfield  {journal} {\bibinfo  {journal} {Phys. Rev. E}\ }\textbf
  {\bibinfo {volume} {102}},\ \bibinfo {pages} {032122} (\bibinfo {year}
  {2020})}\BibitemShut {NoStop}%
\bibitem [{\citenamefont {Papoulis}(1984)}]{Papoulis:BOOK:1984}%
  \BibitemOpen
  \bibfield  {author} {\bibinfo {author} {\bibfnamefont {A.}~\bibnamefont
  {Papoulis}},\ }\href@noop {} {\emph {\bibinfo {title} {{Probability, random
  variables, and stochastic processes}}}},\ \bibinfo {edition} {2nd}\ ed.\
  (\bibinfo  {publisher} {McGraw-Hill},\ \bibinfo {year} {1984})\BibitemShut
  {NoStop}%
\bibitem [{\citenamefont {Wansleben}\ and\ \citenamefont
  {Landau}(1987)}]{Wansleben:JAP:1987}%
  \BibitemOpen
  \bibfield  {author} {\bibinfo {author} {\bibfnamefont {S.}~\bibnamefont
  {Wansleben}}\ and\ \bibinfo {author} {\bibfnamefont {D.~P.}\ \bibnamefont
  {Landau}},\ }\bibfield  {title} {\bibinfo {title} {{Dynamical critical
  exponent of the 3D Ising model}},\ }\href@noop {} {\bibfield  {journal}
  {\bibinfo  {journal} {J. Appl. Phys.}\ }\textbf {\bibinfo {volume} {61}},\
  \bibinfo {pages} {3968} (\bibinfo {year} {1987})}\BibitemShut {NoStop}%
\bibitem [{\citenamefont {Fisher}\ and\ \citenamefont
  {Barber}(1972)}]{Fisher:PRL:1972}%
  \BibitemOpen
  \bibfield  {author} {\bibinfo {author} {\bibfnamefont {M.~E.}\ \bibnamefont
  {Fisher}}\ and\ \bibinfo {author} {\bibfnamefont {M.~N.}\ \bibnamefont
  {Barber}},\ }\bibfield  {title} {\bibinfo {title} {{Scaling Theory for
  Finite-Size Effects in the Critical Region}},\ }\href@noop {} {\bibfield
  {journal} {\bibinfo  {journal} {Phys. Rev. Lett.}\ }\textbf {\bibinfo
  {volume} {28}},\ \bibinfo {pages} {1516} (\bibinfo {year}
  {1972})}\BibitemShut {NoStop}%
\bibitem [{\citenamefont {Botet}\ \emph {et~al.}(1982)\citenamefont {Botet},
  \citenamefont {Jullien},\ and\ \citenamefont {Pfeuty}}]{Botet:PRL:1982}%
  \BibitemOpen
  \bibfield  {author} {\bibinfo {author} {\bibfnamefont {R.}~\bibnamefont
  {Botet}}, \bibinfo {author} {\bibfnamefont {R.}~\bibnamefont {Jullien}},\
  and\ \bibinfo {author} {\bibfnamefont {P.}~\bibnamefont {Pfeuty}},\
  }\bibfield  {title} {\bibinfo {title} {{Size Scaling for Infinitely
  Coordinated Systems}},\ }\href@noop {} {\bibfield  {journal} {\bibinfo
  {journal} {Phys. Rev. Lett.}\ }\textbf {\bibinfo {volume} {49}},\ \bibinfo
  {pages} {478} (\bibinfo {year} {1982})}\BibitemShut {NoStop}%
\bibitem [{\citenamefont {Botet}\ and\ \citenamefont
  {Jullien}(1983)}]{Botet:PRB:1983}%
  \BibitemOpen
  \bibfield  {author} {\bibinfo {author} {\bibfnamefont {R.}~\bibnamefont
  {Botet}}\ and\ \bibinfo {author} {\bibfnamefont {R.}~\bibnamefont
  {Jullien}},\ }\bibfield  {title} {\bibinfo {title} {{Large-size critical
  behavior of infinitely coordinated systems}},\ }\href@noop {} {\bibfield
  {journal} {\bibinfo  {journal} {Phys. Rev. B}\ }\textbf {\bibinfo {volume}
  {28}},\ \bibinfo {pages} {3955} (\bibinfo {year} {1983})}\BibitemShut
  {NoStop}%
\bibitem [{\citenamefont {Colonna-Romano}\ \emph {et~al.}(2014)\citenamefont
  {Colonna-Romano}, \citenamefont {Gould},\ and\ \citenamefont
  {Klein}}]{Romano:PRE:2014}%
  \BibitemOpen
  \bibfield  {author} {\bibinfo {author} {\bibfnamefont {L.}~\bibnamefont
  {Colonna-Romano}}, \bibinfo {author} {\bibfnamefont {H.}~\bibnamefont
  {Gould}},\ and\ \bibinfo {author} {\bibfnamefont {W.}~\bibnamefont {Klein}},\
  }\bibfield  {title} {\bibinfo {title} {{Anomalous mean-field behavior of the
  fully connected Ising model}},\ }\href@noop {} {\bibfield  {journal}
  {\bibinfo  {journal} {Phys. Rev. E}\ }\textbf {\bibinfo {volume} {90}},\
  \bibinfo {pages} {042111} (\bibinfo {year} {2014})}\BibitemShut {NoStop}%
\bibitem [{\citenamefont {Ellis}\ \emph {et~al.}(2010)\citenamefont {Ellis},
  \citenamefont {Machta},\ and\ \citenamefont {Otto}}]{Ellis:AAP:2010}%
  \BibitemOpen
  \bibfield  {author} {\bibinfo {author} {\bibfnamefont {R.~S.}\ \bibnamefont
  {Ellis}}, \bibinfo {author} {\bibfnamefont {J.}~\bibnamefont {Machta}},\ and\
  \bibinfo {author} {\bibfnamefont {P.~T.-H.}\ \bibnamefont {Otto}},\
  }\bibfield  {title} {\bibinfo {title} {Asymptotic behavior of the finite-size
  magnetization as a function of the speed of approach to criticality},\
  }\href@noop {} {\bibfield  {journal} {\bibinfo  {journal} {Ann. Appl.
  Probab.}\ }\textbf {\bibinfo {volume} {20}},\ \bibinfo {pages} {2118 }
  (\bibinfo {year} {2010})}\BibitemShut {NoStop}%
\bibitem [{\citenamefont {Kawashima}\ and\ \citenamefont
  {Ito}(1993)}]{Kawashima:JPSJ:1993}%
  \BibitemOpen
  \bibfield  {author} {\bibinfo {author} {\bibfnamefont {N.}~\bibnamefont
  {Kawashima}}\ and\ \bibinfo {author} {\bibfnamefont {N.}~\bibnamefont
  {Ito}},\ }\bibfield  {title} {\bibinfo {title} {{Critical Behavior of the
  Three-Dimensional $\pm J$ Model in a Magnetic Field}},\ }\href@noop {}
  {\bibfield  {journal} {\bibinfo  {journal} {J. Phys. Soc. Jpn.}\ }\textbf
  {\bibinfo {volume} {62}},\ \bibinfo {pages} {435} (\bibinfo {year}
  {1993})}\BibitemShut {NoStop}%
\bibitem [{\citenamefont {Houdayer}\ and\ \citenamefont
  {Hartmann}(2004)}]{Houdayer:PRB:2004}%
  \BibitemOpen
  \bibfield  {author} {\bibinfo {author} {\bibfnamefont {J.}~\bibnamefont
  {Houdayer}}\ and\ \bibinfo {author} {\bibfnamefont {A.~K.}\ \bibnamefont
  {Hartmann}},\ }\bibfield  {title} {\bibinfo {title} {{Low-temperature
  behavior of two-dimensional Gaussian Ising spin glasses}},\ }\href@noop {}
  {\bibfield  {journal} {\bibinfo  {journal} {Phys. Rev. B}\ }\textbf {\bibinfo
  {volume} {70}},\ \bibinfo {pages} {014418} (\bibinfo {year}
  {2004})}\BibitemShut {NoStop}%
\bibitem [{\citenamefont {Melchert}\ and\ \citenamefont
  {Hartmann}(2009)}]{Melchert:PRB:2009}%
  \BibitemOpen
  \bibfield  {author} {\bibinfo {author} {\bibfnamefont {O.}~\bibnamefont
  {Melchert}}\ and\ \bibinfo {author} {\bibfnamefont {A.~K.}\ \bibnamefont
  {Hartmann}},\ }\bibfield  {title} {\bibinfo {title} {{Scaling behavior of
  domain walls at the $T=0$ ferromagnet to spin-glass transition}},\
  }\href@noop {} {\bibfield  {journal} {\bibinfo  {journal} {Phys. Rev. B}\
  }\textbf {\bibinfo {volume} {79}},\ \bibinfo {pages} {184402} (\bibinfo
  {year} {2009})}\BibitemShut {NoStop}%
\bibitem [{\citenamefont {Melchert}(2014)}]{autoScale:GH:2022}%
  \BibitemOpen
  \bibfield  {author} {\bibinfo {author} {\bibfnamefont {O.}~\bibnamefont
  {Melchert}},\ }\href@noop {} {\bibinfo {title} {{autoScale -- A standalone
  python tool for performing automated finite-size scaling analysis}}},\
  \bibinfo {howpublished} {\url{https://github.com/omelchert/autoScale.git}}
  (\bibinfo {year} {2014})\BibitemShut {NoStop}%
\bibitem [{\citenamefont {Harada}(2011)}]{Harada:PRE:2011}%
  \BibitemOpen
  \bibfield  {author} {\bibinfo {author} {\bibfnamefont {K.}~\bibnamefont
  {Harada}},\ }\bibfield  {title} {\bibinfo {title} {Bayesian inference in the
  scaling analysis of critical phenomena},\ }\href@noop {} {\bibfield
  {journal} {\bibinfo  {journal} {Phys. Rev. E}\ }\textbf {\bibinfo {volume}
  {84}},\ \bibinfo {pages} {056704} (\bibinfo {year} {2011})}\BibitemShut
  {NoStop}%
\bibitem [{\citenamefont {Harada}(2015)}]{Harada:PRE:2015}%
  \BibitemOpen
  \bibfield  {author} {\bibinfo {author} {\bibfnamefont {K.}~\bibnamefont
  {Harada}},\ }\bibfield  {title} {\bibinfo {title} {Kernel method for
  corrections to scaling},\ }\href@noop {} {\bibfield  {journal} {\bibinfo
  {journal} {Phys. Rev. E}\ }\textbf {\bibinfo {volume} {92}},\ \bibinfo
  {pages} {012106} (\bibinfo {year} {2015})}\BibitemShut {NoStop}%
\bibitem [{\citenamefont {Challa}\ \emph {et~al.}(1986)\citenamefont {Challa},
  \citenamefont {Landau},\ and\ \citenamefont {Binder}}]{Challa:PRB:1986}%
  \BibitemOpen
  \bibfield  {author} {\bibinfo {author} {\bibfnamefont {M.~S.~S.}\
  \bibnamefont {Challa}}, \bibinfo {author} {\bibfnamefont {D.~P.}\
  \bibnamefont {Landau}},\ and\ \bibinfo {author} {\bibfnamefont
  {K.}~\bibnamefont {Binder}},\ }\bibfield  {title} {\bibinfo {title}
  {Finite-size effects at temperature-driven first-order transitions},\
  }\href@noop {} {\bibfield  {journal} {\bibinfo  {journal} {Phys. Rev. B}\
  }\textbf {\bibinfo {volume} {34}},\ \bibinfo {pages} {1841} (\bibinfo {year}
  {1986})}\BibitemShut {NoStop}%
\bibitem [{Hou()}]{Houdayer:PRB:2004:S}%
  \BibitemOpen
  \href@noop {} {}\bibinfo {note} {The quality function $S$ measures the
  mean-square distance of the rescaled quantities at finite system size to
  their master curve in units of the standard error. The error $\delta u$ for a
  scaling parameter $u$ is estimated as $S(u \pm \delta u) =
  \mathrm{min}_{u}[S(u)] + 1$.}\BibitemShut {Stop}%
\bibitem [{\citenamefont {Allen}\ and\ \citenamefont
  {Tildesley}(2017)}]{Allen:BOOK:217}%
  \BibitemOpen
  \bibfield  {author} {\bibinfo {author} {\bibfnamefont {M.~P.}\ \bibnamefont
  {Allen}}\ and\ \bibinfo {author} {\bibfnamefont {D.~J.}\ \bibnamefont
  {Tildesley}},\ }\href@noop {} {\emph {\bibinfo {title} {{Computer Simulation
  of Liquids}}}}\ (\bibinfo  {publisher} {Oxford University Press},\ \bibinfo
  {year} {2017})\BibitemShut {NoStop}%
\bibitem [{\citenamefont {Ma}\ \emph {et~al.}(2010)\citenamefont {Ma},
  \citenamefont {Dudarev}, \citenamefont {Semenov},\ and\ \citenamefont
  {Woo}}]{Ma:PRE:2010}%
  \BibitemOpen
  \bibfield  {author} {\bibinfo {author} {\bibfnamefont {P.-W.}\ \bibnamefont
  {Ma}}, \bibinfo {author} {\bibfnamefont {S.~L.}\ \bibnamefont {Dudarev}},
  \bibinfo {author} {\bibfnamefont {A.~A.}\ \bibnamefont {Semenov}},\ and\
  \bibinfo {author} {\bibfnamefont {C.~H.}\ \bibnamefont {Woo}},\ }\bibfield
  {title} {\bibinfo {title} {Temperature for a dynamic spin ensemble},\
  }\href@noop {} {\bibfield  {journal} {\bibinfo  {journal} {Phys. Rev. E}\
  }\textbf {\bibinfo {volume} {82}},\ \bibinfo {pages} {031111} (\bibinfo
  {year} {2010})}\BibitemShut {NoStop}%
\bibitem [{\citenamefont {Ma}\ and\ \citenamefont
  {Dudarev}(2011)}]{Ma:PRB:2011}%
  \BibitemOpen
  \bibfield  {author} {\bibinfo {author} {\bibfnamefont {P.-W.}\ \bibnamefont
  {Ma}}\ and\ \bibinfo {author} {\bibfnamefont {S.~L.}\ \bibnamefont
  {Dudarev}},\ }\bibfield  {title} {\bibinfo {title} {Langevin spin dynamics},\
  }\href@noop {} {\bibfield  {journal} {\bibinfo  {journal} {Phys. Rev. B}\
  }\textbf {\bibinfo {volume} {83}},\ \bibinfo {pages} {134418} (\bibinfo
  {year} {2011})}\BibitemShut {NoStop}%
\bibitem [{\citenamefont {Butler}\ \emph {et~al.}(1998)\citenamefont {Butler},
  \citenamefont {Ayton}, \citenamefont {Jepps},\ and\ \citenamefont
  {Evans}}]{Butler:JCP:1998}%
  \BibitemOpen
  \bibfield  {author} {\bibinfo {author} {\bibfnamefont {B.~D.}\ \bibnamefont
  {Butler}}, \bibinfo {author} {\bibfnamefont {G.}~\bibnamefont {Ayton}},
  \bibinfo {author} {\bibfnamefont {O.~G.}\ \bibnamefont {Jepps}},\ and\
  \bibinfo {author} {\bibfnamefont {D.~J.}\ \bibnamefont {Evans}},\ }\bibfield
  {title} {\bibinfo {title} {{Configurational temperature: Verification of
  Monte Carlo simulations}},\ }\href@noop {} {\bibfield  {journal} {\bibinfo
  {journal} {J. of Chem. Phys.}\ }\textbf {\bibinfo {volume} {109}},\ \bibinfo
  {pages} {6519} (\bibinfo {year} {1998})}\BibitemShut {NoStop}%
\bibitem [{\citenamefont {Rugh}(1997)}]{Rugh:PRL:1997}%
  \BibitemOpen
  \bibfield  {author} {\bibinfo {author} {\bibfnamefont {H.~H.}\ \bibnamefont
  {Rugh}},\ }\bibfield  {title} {\bibinfo {title} {{Dynamical Approach to
  Temperature}},\ }\href@noop {} {\bibfield  {journal} {\bibinfo  {journal}
  {Phys. Rev. Lett.}\ }\textbf {\bibinfo {volume} {78}},\ \bibinfo {pages}
  {772} (\bibinfo {year} {1997})}\BibitemShut {NoStop}%
\bibitem [{\citenamefont {Rugh}(1998)}]{Rugh:JPA:1998}%
  \BibitemOpen
  \bibfield  {author} {\bibinfo {author} {\bibfnamefont {H.~H.}\ \bibnamefont
  {Rugh}},\ }\bibfield  {title} {\bibinfo {title} {A geometric, dynamical
  approach to thermodynamics},\ }\href@noop {} {\bibfield  {journal} {\bibinfo
  {journal} {J. Phys. A: Math. Gen.}\ }\textbf {\bibinfo {volume} {31}},\
  \bibinfo {pages} {7761} (\bibinfo {year} {1998})}\BibitemShut {NoStop}%
\bibitem [{\citenamefont {Nurdin}\ and\ \citenamefont
  {Schotte}(2002)}]{Nurdin:PA:2002}%
  \BibitemOpen
  \bibfield  {author} {\bibinfo {author} {\bibfnamefont {W.~B.}\ \bibnamefont
  {Nurdin}}\ and\ \bibinfo {author} {\bibfnamefont {K.-D.}\ \bibnamefont
  {Schotte}},\ }\bibfield  {title} {\bibinfo {title} {{Dynamical temperature
  study for classical planar spin systems}},\ }\href@noop {} {\bibfield
  {journal} {\bibinfo  {journal} {Physica A}\ }\textbf {\bibinfo {volume}
  {308}},\ \bibinfo {pages} {209} (\bibinfo {year} {2002})}\BibitemShut
  {NoStop}%
\bibitem [{\citenamefont {Ota}\ \emph {et~al.}(1992)\citenamefont {Ota},
  \citenamefont {Ota},\ and\ \citenamefont {F\"ahnle}}]{Ota:JPC:1992}%
  \BibitemOpen
  \bibfield  {author} {\bibinfo {author} {\bibfnamefont {S.}~\bibnamefont
  {Ota}}, \bibinfo {author} {\bibfnamefont {S.~B.}\ \bibnamefont {Ota}},\ and\
  \bibinfo {author} {\bibfnamefont {M.}~\bibnamefont {F\"ahnle}},\ }\bibfield
  {title} {\bibinfo {title} {{Microcanonical Monte Carlo simulations for the
  two-dimensional XY model}},\ }\href@noop {} {\bibfield  {journal} {\bibinfo
  {journal} {J. Phys.: Condens. Matter}\ }\textbf {\bibinfo {volume} {4}},\
  \bibinfo {pages} {5411} (\bibinfo {year} {1992})}\BibitemShut {NoStop}%
\bibitem [{\citenamefont {Bollob\'as}(2001)}]{Bollobas:BOOK:2001}%
  \BibitemOpen
  \bibfield  {author} {\bibinfo {author} {\bibfnamefont {B.}~\bibnamefont
  {Bollob\'as}},\ }\href@noop {} {\emph {\bibinfo {title} {{Random Graphs}}}},\
  \bibinfo {edition} {2nd}\ ed.\ (\bibinfo  {publisher} {Cambridge University
  Press},\ \bibinfo {year} {2001})\BibitemShut {NoStop}%
\bibitem [{\citenamefont {Melchert}\ \emph {et~al.}(2011)\citenamefont
  {Melchert}, \citenamefont {Hartmann},\ and\ \citenamefont
  {M\'ezard}}]{Melchert:PRE:2011}%
  \BibitemOpen
  \bibfield  {author} {\bibinfo {author} {\bibfnamefont {O.}~\bibnamefont
  {Melchert}}, \bibinfo {author} {\bibfnamefont {A.~K.}\ \bibnamefont
  {Hartmann}},\ and\ \bibinfo {author} {\bibfnamefont {M.}~\bibnamefont
  {M\'ezard}},\ }\bibfield  {title} {\bibinfo {title} {Mean-field behavior of
  the negative-weight percolation model on random regular graphs},\ }\href@noop
  {} {\bibfield  {journal} {\bibinfo  {journal} {Phys. Rev. E}\ }\textbf
  {\bibinfo {volume} {84}},\ \bibinfo {pages} {041106} (\bibinfo {year}
  {2011})}\BibitemShut {NoStop}%
\end{thebibliography}%

\end{document}